\documentclass[10pt]{article}

\usepackage{amsmath}
\usepackage{graphicx,psfrag,epsf}
\usepackage{enumerate}
\usepackage{natbib}
\usepackage{url} % not crucial - just used below for the URL
\usepackage{amssymb}
\usepackage{array}
\usepackage{multirow}
\usepackage{amsthm}
\usepackage[table]{xcolor}
\definecolor{darkblue}{rgb}{0.0,0.0,0.5}
\usepackage{comment}
\usepackage{marginnote}
\newcommand{\iid}{\stackrel{iid}{\sim}}

\newcommand{\ud}{\text{d}}
\newcommand{\tbz}{\widetilde{\bz}_\emsize}
\newcommand{\fit}{\text{\tiny fit}}

\newcommand{\Normal}{\mathcal{N}}
\newcommand{\Var}{\text{Var}}

\newcommand{\bz}{\boldsymbol\zeta}

\newcommand{\mix}{\text{\tiny mix}}
\newcommand{\pmix}{\text{\tiny (mix)}}
\newcommand{\pU}{\text{\tiny (U)}}
\newcommand{\pX}{\text{\tiny ($\mathcal{X}$)}}

\newcommand{\pI}{\text{\tiny (I)}}
\newcommand{\pII}{\text{\tiny (II)}}
\newcommand{\pIII}{\text{\tiny (III)}}
\newcommand{\ttEM}{\text{\tiny EM}}
\newcommand{\ttBS}{\text{\tiny BS}}
\newcommand{\tr}{\mathcal{F}}
\newcommand{\trh}{\mathcal{H}}
\newcommand{\scale}{\mathcal{S}}
\newcommand{\E}{\text{E}}
\newcommand{\pk}{\text{\tiny$(k)$}}
\newcommand{\imp}{\text{\tiny imp}}
\newcommand{\geo}{\text{\tiny geo}}
\newcommand{\opt}{\text{\tiny opt}}
\newcommand{\emsize}{\text{\tiny $L$}}

\newcommand{\LL}{\text{\tiny $L$}}
\newcommand{\measure}{\mathbf{u}}
\newcommand{\measurev}{\mathbf{v}}

\RequirePackage[colorlinks,allcolors=darkblue]{hyperref}

\newcommand{\blind}{1}

\begin{document}

\def\spacingset#1{\renewcommand{\baselinestretch}%
{#1}\small\normalsize} \spacingset{1}

%%%%%%%%%%%%%%%%%%%%%%%%%%%%%%%%%%%%%%%%%%%%%%%%%%%%%%%%%%%%%%%%%%%%%%%%%%%%%%

\if0\blind
{
  \title{\bf Warp Bridge Sampling:  The Next Generation}
  \author{BLIND VERSION\thanks{BLIND VERSION } \hspace{.2cm}\\
    \\
%    and \\
%    Author 2 \\
%    Department of ZZZ, University of WWW
    }
  \maketitle
} \fi

\if1\blind
{
  \title{\bf Warp Bridge Sampling: The Next Generation}
  \author{Lazhi Wang$^1$, David E. Jones$^2$, and Xiao-Li Meng$^3$\thanks{The authors gratefully
acknowledge helpful conversations with members of the Department of Statistics at Harvard University, constructive comments from the audience of the 2016 MCQMC conference at Stanford University, and
partial financial support from NSF and JTF. For correspondence, email david.jones@tamu.edu.}\\\\\hspace{.2cm}Two Sigma Investments, LP$^1$\thanks{The views expressed herein
 are solely the views of the author(s) and are not necessarily the views of Two Sigma Investments, LP or any of its affiliates.  They are not intended to provide,
 and should not be relied upon for, investment advice. Please see the full disclaimer on page 30.}, Texas A\&M University$^2$, and Harvard University$^3$\hspace{.2cm}\\\\
    }
  \maketitle
} \fi
\kern -.2in
%\spacingset{1.45}
\begin{abstract}
\noindent Bridge sampling   is an effective Monte Carlo  method for estimating the ratio of normalizing constants of two probability densities,   a  routine  computational problem  in statistics, physics, chemistry,  and   other fields. The Monte Carlo error of the bridge sampling estimator is determined  by the amount of overlap between the two densities. In the case of uni-modal densities, Warp-I, II, and III transformations  \citep{meng2002warp} are effective for increasing the initial overlap, but they are less so for multi-modal densities. This paper introduces Warp-U transformations that aim to transform multi-modal densities into \textit{Uni-modal} ones without altering their normalizing constants. The construction of a Warp-U transformation starts with  a Normal (or other convenient)  mixture distribution $\phi_{\mix}$ that has reasonable overlap with the target density $p$, whose normalizing constant is unknown.  The stochastic transformation that maps $\phi_{\mix}$ back to its generating distribution  $\Normal(0,1)$ is then applied to  $p$ yielding its Warp-U version, which we denote $\tilde p$. Typically, $\tilde p$ is uni-modal and has substantially increased overlap with $\phi$. Furthermore, we prove that the overlap between $\tilde p$ and $\Normal(0,1)$ is guaranteed to be no less than the overlap between $p$ and  $\phi_{\mix}$, in terms of any $f$-divergence. %In practice, the overlap between $\tilde p$ and $\Normal(0,1)$ is almost always larger.
%Consequently, Warp-U transformations usually lead to statistically more efficient  bridge sampling estimators, and never reduce statistical efficiency.
We propose a computationally efficient method to find an appropriate  $\phi_{\mix}$,
and a simple but effective approach to remove the bias which results from estimating the normalizing constants and fitting $\phi_{\mix}$ with the same data.
%{\color{red} and  use simulations  to explore various estimation strategies and choices for tuning parameters, with the aim to achieve statistical efficiency gains without unduly losing computational efficiency}.
We illustrate our findings using 10 and 50 dimensional highly irregular multi-modal densities, and demonstrate how Warp-U sampling can be used to improve the final estimation step of the Generalized Wang-Landau algorithm \citep{liang2005generalized}, a powerful sampling and estimation method.
\end{abstract}

\noindent%

%\newpage
\spacingset{1.45} % DON'T change the spacing!

%%%%%%%%%%%%%%%%%%%%%%%%%%%%%%%%%%%%%%%%%%%%%%%%%%%%%%
\section{Motivation}\label{sec:introduction}

Markov chain Monte Carlo (MCMC)  methods, such as the Metropolis-Hastings algorithm,  enable us to simulate from an unnormalized density without
knowing its normalizing constant. However, in many scientific and statistical
studies the very quantities of interest are normalizing constants or ratios of them; see, for example,  \citet{voter1985monte}, \citet{kass1995bayes}, \citet{meng1996simulating}, \citet{diciccio1997computing}, \citet{gelman1998simulating}, \citet{shao2000monte}, and \citet{tan2013calibrated}. A well-known example from physics and chemistry is the computation of a partition function,
which describes the statistical properties of a system in thermodynamic equilibrium.
A partition function
is the integral of an unnormalized
density $q(\omega;T,v)=\exp\left\{-H(\omega,v)/(kT)\right\},$
where $T$ is  temperature, $k$ is Boltzmann's constant, $v$ is a vector of
system characteristics, and $H(\omega, v)$ is the energy function. Because of the
high dimensionality of $H$, Monte Carlo (MC) methods are often the only feasible  tool for
estimating  a partition function,  i.e., the normalizing constant of $q;$ see \cite{bennett1976efficient}, \cite{voter1985dynamical}, and \cite{ceperley1995path}.

Two key objects in statistics which can be expressed as  normalizing constants are observed-data likelihoods and  Bayes factors. Focusing on the latter,
%Another example is the computation of the observed-data likelihood, $L(\Theta;Y_{\obs})$, which is the normalizing constant of the conditional distribution of  $Y_\mis$ given $(Y_\obs,\Theta)$, $P(Y_\obs|Y_\mis,\Theta)$, with the complete-data distribution as an unnormalized density,
%the complete-data likelihood, %$P(y_{obs},y_{mis}|\Theta)$
%i.e., \begin{equation*} L(\Theta;Y_{\obs}) \triangleq P(Y_{\obs}|\Theta) = \int P(Y_{\mis},Y_{\obs}|\Theta)\measure(\ud Y_{\mis}),\end{equation*} where $\measure$ is the underlying measure (e.g., Lebesgue or a counting measure). As an example in genetics,  $\Theta$ represents the locations of disease genes relative to a set of markers, $Y_{\obs}$ is a vector of genotypes of markers for some members of a pedigree, and $Y_{\mis}$ represents unobserved allele  types inherited from parents. For a large pedigree with many loci, direct calculation of the observed-data likelihood is often prohibitive. Fortunately, it is feasible to evaluate $P(Y_{\obs}, Y_{\mis}|\Theta)$ and to simulate $Y_{\mis}$ from the conditional distribution, $P(Y_{\mis}|Y_{\obs},\Theta)$; therefore estimating $L(\Theta;Y_{\obs})$ becomes estimating the normalizing constant of $P(Y_{\mis}|Y_{\obs}, \Theta) \propto P(Y_{\mis},Y_{\obs}|\Theta).$
%In addition, MC integration is often used to estimate Bayes factors. Specifically,
let $Y$ be our data, and let $M_0$ and $M_1$ be two plausible models parameterized by
$\Theta_0$ and $\Theta_1$, respectively. The Bayes factor is then  the ratio of
the model likelihoods, $P(Y|M_0)$ and $P(Y|M_1)$, where
\begin{equation*}
P(Y|M_i)=\int P(Y|\Theta_i,M_i)P(\Theta_i|M_i)\measure(\ud\Theta_i)
\end{equation*}
%is the probability of $Y$ integrated over
%the prior distribution of $\Theta_i$ under $M_i$. $P(Y|M_i)$
is  the normalizing
constant of the unnormalized density, $P(\Theta_i,Y|M_i)\propto P(Y|\Theta_i,M_i)P(\Theta_i|M_i)$, for $i=1,2$. %, of $\Theta_i$.
In most Bayesian analyses, MC draws of $\Theta_i$ from
$P(\Theta_i|Y,M_i)$ are made for the purpose of statistical
inference, often using Monte Carlo Markov chain (MCMC) methods. Hence, to estimate  $P(Y|M_i)$, it is desirable to use methods that require only  these available draws (plus perhaps some draws from another convenient distribution).

One such method is the bridge sampling approach introduced by \citet{bennett1976efficient} and generalized and popularized by \citet{meng1996simulating}. In this paper we propose an approach to improve the efficiency of bridge sampling in the multi-modal context, which is common when using even moderately complex models. Specifically, we introduce a class of stochastic transformations, Warp-U transformations,
that can warp two multi-modal densities into densities having substantial overlap but without changing their respective normalizing constants. For bridge sampling, an increase in distributional overlap implies superior statistical efficiency. The key idea of Warp-U transformations is to approximate the unnormalized density of interest $q$ by a mixture distribution (e.g., a Normal mixture), and then to construct a coupling which allows us to (stochastically) map $q$ into a uni-modal density in the same way that the approximating mixture can be mapped back to a single generative density (e.g., a single Normal density). Our work builds on the warp transformations (centering, scaling, and symmetrizing) for uni-modal densities that were proposed by \citet{meng2002warp}.  Our method also has an intrinsic connection with the theory of optimal transport \citep[e.g.][]{villani2003topics}, albeit here we typically only seek a reasonable transport (from one density to another) which can achieve a beneficial compromise between statistical efficiency and computational efficiency.  %These earlier transformations (and stochastic transformations) involve centering, scaling, and symmetrizing the densities in question. %, so that they have similar shapes and locations. %Warp-U transformations are necessarily somewhat more complex because in the case of a multi-modal density we must first identify the individual components and then transform them together or one at a time.

The utility of Warp-U transformations  is especially promising because bridge sampling is similar to many other mixture sampling approaches. Indeed, although developed from different perspectives,  a number of methods in the literature turn out to be special cases of bridge sampling or adaptations of it, as demonstrated by \cite{mira2004bridge}. For instance, the marginal likelihood approaches of \citet{chib1995marginal} and \citet{chib2001marginal} based on Metropolis-Hastings output correspond to bridge sampling with specific choices of the bridge density.  Similarly,  the defensive sampling method of \citet{hesterberg1995weighted} for estimating normalizing constants can be directly interpreted as bridge sampling. The `balance weight heuristic' introduced by \citet{veach1995optimally} is a generalization of bridge sampling where the unnormalized densities to be integrated are not necessarily included among the sampling densities. This more general bridge sampling is also covered by the likelihood approach proposed by \citet{kong2003theory}, which reformulates MC integration as an estimation problem with the dominating measure as the estimand. %, and identifies the maximum likelihood estimator which is a finite   %; their idea is to formalize the approximation  %The likelihood approach is natural since the key idea of MC  integration is to approximate a integral with respect to an intractable/innumerable measure (e.g., the Lebesgue measure) by an integral with respect to a finitely supported measure, i.e., a finite sum.
Their likelihood framework provides a unified way of deriving and characterizing various strategies for boosting statistical   efficiency, such as the control variates method of \citet{owen2000safe}; see \citet{tan2004likelihood}, \citet{MengXiao-Li2005CCSa}, and  \citet{kong2006} for details and illustrations. As
discussed in Section \ref{sec:base_and_mle}, there is a possibility to
cast Warp-U bridge sampling methods under the same likelihood framework and thereby make further connections with other methods, but that is a topic for future exploration.

Warp-U transformations can additionally be useful in contexts where bridge sampling or analogous approaches are not applied or not initially applied. For example, %Following a more traditional way of constructing MC estimates,
\citet{elvira2015efficient} proposed a mixture sampling approach that, instead of using a mixture of all the sampling densities as MC weights, aims to save computation by using only some of the sampling densities to compute each weight. This is statistically less efficient than bridge sampling, but is based on the same ideas, and would also benefit from greater overlap between the sampling and target densities. %Furthermore, Warp-U transformations could well prove to be useful in other areas of Bayesian computation since multi-modal densities are a ubiquitous challenge.
For another example, consider the
%There are additionally a number of
powerful adaptive importance sampling based methods which can be used to estimate normalizing constants, e.g., the multicanonical algorithm \citep{berg1991multicanonical}, the $1/k$-ensemble algorithm \citep{hesselbo1995monte}, dynamic weighting \citep[e.g.,][]{wong1997dynamic,liu2001theory}, the Wang-Landau algorithm and adaptations of it \citep[e.g.,][]{wang2001efficient,liang2005generalized,liang2007stochastic,bornn2013adaptive}, and layered adaptive importance sampling \citep{martino2017layered}. These are sampling methods, in contrast to (Warp-U) bridge sampling,
which is a \textit{post-sampling} method.  Therefore, they
are not directly comparable (e.g., costs for sampling and the final estimation  are typically rather different).  %, it does not say anything about how to obtain these samples or quantify the computational cost of doing so, and any comparison would clearly need to account for these costs.
Instead, we show how Warp-U bridge sampling can be used as an additional step to improve adaptive importance sampling algorithms. Specifically, we illustrate that target draws can be obtained by re-sampling from the output of the Generalized Wang-Landau (GWL) algorithm \citep{liang2005generalized}, and that applying Warp-U bridge sampling to these draws  can substantially reduce the mean squared error (MSE) of the final normalizing constant estimators. %Our approach is a straightforward extension of the GWL algorithm and we anticipate that it can be applied with only minor alterations to other adaptive importance sampling approaches.

%The goal of warp-U transformations are
%The goal of warp-U transformations, and warp transformation generally, is similar to There are also related techniques  improve importance sampling, but in a somewhat different way to how warp transformations improve bridge sampling.Since the methods use different information we do not compare them here.

A limitation of our work is that when we take into account computational cost Warp-U bridge sampling is not necessarily superior to the following approach: first compute a mixture approximation to the target density exactly as in Warp-U bridge sampling, and then apply ordinary bridge sampling making use of this approximate density (see Section \ref{subsection::simulation_known_mixture}). However, Warp-U bridge sampling is a valuable development in that it fundamentally differs from this ordinary method and therefore provides new opportunities for improving computational efficiency in terms of precision per second. In particular, we anticipate that further work will yield approximations to Warp-U bridge sampling that enable it to surpass simpler methods based on ordinary bridge sampling, especially because here we have made no attempt to optimize our code or implementation; see the discussion in Section \ref{sec:discussion}.

Our paper is organized as follows. Section 2 briefly overviews bridge sampling and the
warp transformations of \citet{meng2002warp}, highlighting their power for increasing distribution overlap.     Section 3  defines and illustrates the Warp-U transformation we propose and then establishes the theoretical result that Warp-U transformations never reduce distribution overlap. Section
\ref{specific_method}  outlines  a computationally efficient strategy  for  finding  a specific Warp-U transformation
 and studies  the properties of the corresponding estimator. Section 5 demonstrates that the estimation performance of the Generalized Wang-Landau (GWL) algorithm introduced by \citet{liang2005generalized} can be improved by combining it with Warp-U bridge sampling.  %\ref{section::Complete_Simulation_Study} compares both the computational cost and the statistical  efficiency of estimators with different tuning parameters, aiming to provide practical guidance for choosing them.
Discussion is found in Section \ref{sec:discussion}. The appendices in the online supplementary material provide a proof of Theorem 1,  %(Section \ref{theorem})},
computational details, and guidance on selecting the tuning parameters of our method. %, e.g., the number of components for the approximating mixture.
%investigates the impact of different expand on our investigations of computational efficiency in Section 6.}

%%%%%%%%%%%%%%%%%%%%%%%%%%%%%%%%%%%%%%%%%%%%%%%%%%%%%%
%\kern -.3in
\section{The Basics of Warp Bridge Sampling}
%\kern -.1in
Bridge sampling \citep{bennett1976efficient,meng1996simulating} estimates
the ratio of the normalizing constants of two unnormalized densities by leveraging the overlap between the two densities.
Any method that can increase this overlap has the potential to reduce the MC error. The warp bridge
sampling of \cite{meng2002warp} explored this idea  by transforming the original MC draws so that the densities of the transformed draws have substantially more overlap.

Let $q_i$ be an unnormalized density with normalizing constant $c_i$, for $i=1,2$. Furthermore, let $\measure$ be the underlying measure common to both densities, typically the Lebesgue or counting measure. We use $p_i$ to denote
the normalized density, i.e.,  $p_i(\omega)=c_i^{-1}q_i(\omega)$, for $\omega\in\Omega_i$, where
$\Omega_i$ is the support of $q_i$. Our goal is to estimate
the ratio
$r=c_1/c_2$ or $\lambda=\log(r)$, using  the  draws, $\{w_{i,1}, w_{i,2},
\ldots, w_{i,n_i}\}$,  from $p_i$, for $i=1,2$.  In some instances, we only wish to estimate one normalizing  constant $c_1$, in which case we will select $q_2 = p_2$ to be a convenient density with $c_2=1$ (discussed in Section \ref{sec:warpu}).
Below, we begin  by assuming that draws from $p_1$ and $p_2$ are given, but in Section \ref{sec:comparison} we combine our estimation strategy  with the GWL sampling algorithm.
%It is beyond the scope of this paper when the draws are not independent.
%However, the results here do provide a
%%%%%%%%%%%%%%%%%%%%%%%%%%%%%%%%%%%%%%%%%%%%%%

%\kern -.2in
\subsection{Bridge Sampling}\label{sec:bridge_sampling}

Here we review the key aspects of bridge sampling that will be used in this paper. For a complete treatment, the reader is referred to \citet{meng1996simulating} and the practical introduction by \citet{gronau2017tutorial}. An \texttt{R} package developed by \citet{gronau2017bridgesampling}  is available at
\url{https://cran.r-project.org/web/packages/bridgesampling/}.

Bridge sampling relies on the fact that for any function, $\alpha$,  defined on $\Omega_1\cap\Omega_2$ and
satisfying
$
0<\left|\int_{\Omega_1\cap\Omega_2} \alpha(\omega)p_1(\omega)p_2(\omega)\measure(\ud\omega)\right|<\infty,
$
the following identity holds;
\begin{equation}\label{eq::bridgesampling}
r=\frac{c_1}{c_2}=\frac{\E_2[q_1(\omega)\alpha(\omega)]}{\E_1[q_2(\omega)
\alpha(\omega)]},
\end{equation}
where $\E_i$ represents expectation with respect to  $p_i$. Here $\alpha$ serves as a ``bridge" connecting $p_1$ and $p_2$.
The bridge sampling estimator of $r$ is the sample counterpart of  (\ref{eq::bridgesampling}), i.e.,
\begin{equation}\label{general_bridge_sampling}
\hat{r}_{\alpha}=\dfrac{n_2^{-1}\displaystyle\sum_{j=1}^{n_2}q_1(w_{2,j})\alpha(w_{2,j})}
{n_1^{-1}\displaystyle\sum_{j=1}^{n_1}q_2(w_{1,j})\alpha(w_{1,j})}.
\end{equation}
For example, both importance sampling and  geometric bridge sampling are special cases of
bridge sampling, with $\alpha_\imp\propto1/q_2$ and
$\alpha_\geo\propto1/\sqrt{q_1q_2}$, respectively.

Different choices of $\alpha$ lead to estimators  with different
statistical efficiency, which we quantify
 by the asymptotic variance of $\hat{\lambda}_{\alpha}=\log(\hat{r}_{\alpha})$, or equivalently,
the asymptotic relative variance of
$\hat{r}_{\alpha}$, $\E(\hat{r}_{\alpha}-r)^2/r^2$.
%Under the assumption that all the MC draws used in (\ref{general_bridge_sampling})
%are mutually independent, \citet{meng1996simulating} showed that the first-order asymptotic variance of $\hat{\lambda}_{\alpha}$ is given by $(n_1+n_2)^{-1}\mathcal{V}_\alpha(p_1,p_2)$, where
%\begin{equation}\label{general_D}\mathcal{V}_\alpha(p_1,p_2) =\frac{\int p_1^*p_2^*(p_1^*+p_2^*)\alpha^2\measure(\ud\omega)}{\left(\int p_1^*p_2^*\alpha\measure(\ud\omega)\right)^2} -\frac{1}{s_1}-\frac{1}{s_2},\end{equation}
%with $s_i=n_i/(n_1+n_2)$ and $p_i^* = s_ip_i$.
%When $\alpha(\omega)\propto(q_2(\omega))^{-1}$, $\hat{r}_{\alpha}$ is the importance
%sampling estimator; when
%$\alpha(\omega)\propto\left(q_1(\omega)q_2(\omega)\right)^{-1/2}$, $\hat{r}_{\alpha}$
%corresponds to the geometric bridge sampling estimator, denoted as $\hat{r}_\geo$.
Under the assumption that all the MC draws used in (\ref{general_bridge_sampling})
are i.i.d. (identically and independently distributed), \citet{meng1996simulating} derived the first-order asymptotic variance of $\hat{\lambda}_{\alpha}$,
from which they found that the optimal bridge has the form
\begin{equation}\label{eq:opti}
\alpha_\opt(\omega)\propto\frac{1}{s_1q_1(\omega)+rs_2q_2(\omega)},\quad
{\rm where} \  s_i=\frac{n_i}{n},\  i=1,2.
\end{equation}

Before we proceed, we emphasize that the bridge sampling method itself does
not require the assumption of i.i.d sampling; otherwise the method would be too limited to deserve a general \texttt{R} package.  The i.i.d. assumption was invoked by \cite{meng1996simulating} to make the theoretical calculation both feasible and insightful, in the sense that the resulting optimal bridge (\ref{eq:opti}) takes an appealing mixture form which provides practical guidance. Indeed, regardless of whether  the i.i.d. assumption holds, (\ref{eq:opti}) provides a very effective bridge. % (for example, it eliminates the instability of importance sampling weights).
In contrast,  without the i.i.d. assumption the optimal bridge has a very involved expression \citep{romero2003two}, and offers little practical guidance. Therefore, for the rest of the article we invoke the i.i.d. assumption only for theoretical claims (e.g., when we refer to the ``optimal" approach) or for simulation simplicity.

%showed that the asymptotic relative variance of the geometric  estimator, $\hat{r}_\geo$, is
%\begin{equation}\label{geometric_variance} \Var\left(\hat{\lambda}_\geo\right) = \left(\frac{1}{n_1}+\frac{1}{n_2}\right) \left\{b\left[1-H_{\text{\tiny E}}^2(p_1,p_2)\right]^{-2}-1\right\} + o\left(\frac{1}{n_1+n_2}\right),\end{equation}
%where $\hat{\lambda}_\geo = \log\left(\hat{r}_\geo\right)$, $b=\int_{\Omega_1\cap\Omega_2}\left[p^*_1(\omega)+p^*_2(\omega)\right]\measure(\ud\omega)\leqslant 1$, and $H_\text{\tiny E}(p_1,p_2)$ is the Hellinger distance between $p_1$ and $p_2$, defined as
%\begin{equation}\label{hellinger_distance} H_\text{\tiny E}(p_1, p_2) = \left[\frac{1}{2}\int\left(\sqrt{p_1(\omega)}-\sqrt{p_2(\omega)}\right)^2\measure(\ud\omega)\right]^{1/2}. \end{equation}
%When all the draws are independent,
%\citet{meng1996simulating} found that

Because  $\alpha_\opt$
depends on the unknown quantity $r$, \citet{meng1996simulating} proposed an iterative sequence that rapidly converges to   $\hat{r}_{\opt}$, i.e.,
\begin{equation}\label{optestimator}
\hat{r}_\opt^{(t+1)}=\frac{\dfrac{1}{n_2}\displaystyle\sum_{j=1}^{n_2}\left[\dfrac{l_{2,j}}{s_1l_{2,j}
+s_2\hat{r}_\opt^{(t)}}\right]}
{\dfrac{1}{n_1}\displaystyle\sum_{j=1}^{n_1}\left[\dfrac{1}{s_1l_{1,j}+s_2\hat{r}_\opt^{(t)}}\right]
},
\end{equation}
where $l_{i,j}=q_1(w_{i,j})/q_2(w_{i,j}), \text{ for } i=1, 2,$ and $ j=1, 2, \ldots, n_i$.
%The sequence typically converges to $\hat{r}_\opt$ within a few  iterations (e.g., 10).
%an iterative sequence is required to obtain $\hat{r}_{\alpha_\opt}$ (for simplicity, we write $\hat{r}_\opt$). \citet{meng1996simulating} proposes the following sequence that converges very quickly (usually within 5 iterations) to $\hat{r}_\opt$:
%\begin{equation}
%\hat{r}_\opt^{(t+1)}=\frac{\dfrac{1}{n_2}\sum_{j=1}^{n_2}\left[\dfrac{l_{2,j}}{s_1l_{2,j}+s_2\hat{r}_\opt^{(t)}}\right]}
%{\dfrac{1}{n_1}\sum_{j=1}^{n_1}\left[\dfrac{1}{s_1l_{1,j}+s_2\hat{r}_\opt^{(t)}}\right]
%},
%\end{equation}
%\cite{meng1996simulating} showed that t
\cite{meng1996simulating} showed that, under the i.i.d. assumption,  the asymptotic variance of $\hat{\lambda}_\opt=\log(\hat{r}_\opt)$ is
\begin{equation}\label{opt_variance}
\left(\frac{1}{n_1}+\frac{1}{n_2}\right)
\left[\left(1-H_\text{\tiny A}(p_1, p_2)\right)^{-1}-1 \right]+o\left(\frac{1}
{n_1+n_2}\right),
\end{equation}
which is the same as the asymptotic variance of  the unobtainable optimal estimator  $\hat{\lambda}_{\alpha_\opt}=\log(\hat r_{\alpha_{\rm opt}})$.
Here $H_\text{\tiny A}(p_1, p_2)$ is the sample-size adjusted harmonic divergence  between $p_1$ and $p_2$: \begin{equation}\label{harmonic_distance}
H_\text{\tiny A}(p_1, p_2)=1 -\int_{\Omega_1\cap\Omega_2} \left[w_1p_1^{-1}(\omega)+w_2p_2^{-1}(\omega)\right]^{-1}\measure(\ud
\omega),\end{equation}with $w_i=s_i^{-1}/(s_1^{-1}+s_2^{-1}), i=1,2.$ Using a likelihood that treats  the baseline measure $\measure$ as the (infinite dimensional) parameter, \cite{kong2003theory} showed that $\hat{r}_\opt$ is the maximum likelihood estimator for $r$ (again, under the i.i.d. assumption), thereby further confirming its optimality.

%\kern -.4in
\subsection{Warp Bridge Sampling}\label{sec:warpbridge}
For $i=1,2$, consider a transformation $\tr_i$ of $w_{i,j}$
%$\{w_{i,1},w_{i,2},\ldots,w_{i,n_i}\}$,
such that (a) the unnormalized density, $\tilde{q}_i$, of the transformed draws, $\tilde{w}_{i,j}=\tr_i(w_{i,j})$, has the
same normalizing constant as $q_i$, and (b)
 $H_{\text{\tiny A}}(\tilde{p}_1,\tilde{p}_2)< H_{\text{\tiny A}}(p_1,p_2)$. Then by (\ref{opt_variance}), the optimal bridge sampling estimator based on the transformed draws $\{(\tilde{w}_{i,1},\ldots,\tilde{w}_{i,n_i}); i=1,2\}$
will have smaller asymptotic variance than that based on the original draws $\{(w_{i,1},\ldots,w_{i,n_i});i=1,2\}$, assuming the draws are independent.
This observation motivated the Warp transformations proposed by \cite{meng2002warp}, whose contribution also demonstrated empirically the benefit of Warp transformations under general MCMC settings (i.e., without requiring i.i.d. draws).

The simple idea of Warp-I transformations is to increase overlap among densities (e.g., in terms of $H_A$ in (\ref{harmonic_distance})) by shifting them so that they share a common location.  Specifically, let $\mu_i$ be a location parameter (e.g., mean or mode) of $p_i$, for $i=1, 2$, and suppose that the dominating measure (e.g., the Lebesgue measure) is invariant to translation.  Let
$\tilde{w}^{\pI}_{i,j}=w_{i,j}-\mu_i$ and denote the corresponding unnormalized density by $\tilde{q}^{\pI}_i(w)=q_i(w+\mu_i)$; clearly this density has the same normalizing constant $c_i$ as the original target $q_i(w)$, for $i=1,2.$
The Warp-I bridge sampling estimator
is then obtained by replacing $w_{i,j}$ and $q_i$ in (\ref{general_bridge_sampling}) with
$\tilde{w}^\pI_{i,j}$ and $\tilde{q}^\pI_i$, respectively.
%\begin{equation*}
%\hat{r}_{\alpha}^{\pI}=\dfrac{n_2^{-1}\sum_{j=1}^{n_2}\tilde{q}^{\pI}_1(\tilde{w}^{\pI}_{2,j})\alpha(\tilde{w}^{\pI}_{2,j})}
%{n_1^{-1}\sum_{j=1}^{n_1}\tilde{q}^{\pI}_2(\tilde{w}^{\pI}_{1,j})\alpha(\tilde{w}^{\pI}_{1,j})}
%=\dfrac{n_2^{-1}\sum_{j=1}^{n_2}q_1(w_{2,j}+\mu)\alpha(w_{2,j})}{n_1^{-1}\sum_{j=1}^{n_1}q_2(w_{1,j}-\mu)\alpha(w_{1,j}-\mu)}.
%\end{equation*}
%Figure \ref{warp1} shows the densities before (left panel) and after (right panel) Warp-I transformation, demonstrating the substantially increased overlap.

% \begin{figure}[t]
% \begin{center}
% \includegraphics[width=0.8\textwidth]{fig1.pdf}
% %\caption{Graphical illustration of Warp-I transformation.}
% \caption{\small  {Graphical illustration of Warp-I transformation.
% The dashed and the solid lines are the curves of $p_1$ and $p_2$.
% The dash-dot line is
% the density, $\tilde{p}^\pI_1$, of the Warp-I transformed samples, from  moving $p_1$ to the left by
% $\mu$ units.
% The shaded areas are the overlap between two densities. }}
% \label{warp1}
% \end{center}\end{figure}

\begin{figure}[t]
\begin{center}
\includegraphics[width=0.8\textwidth]{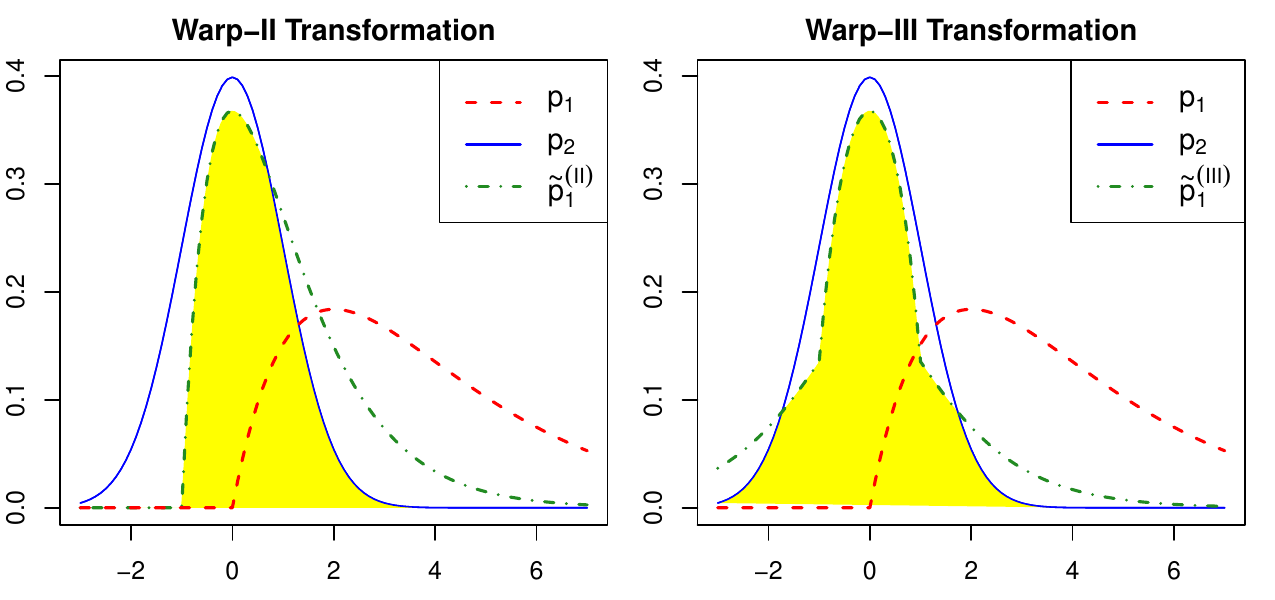}
%\caption{Graphical illustration of Warp-II and Warp-III transformation.}
\caption{\label{warp2}\small {Graphical illustration of Warp-II (left) and Warp-III transformations (right).
The dashed and the solid lines are the curves of $p_1$ and $p_2$.
The dash-dot lines are $p_1^\pII$ (left) and
$p_1^\pIII$ (right), obtained by Warp-II and Warp-III transformation, respectively.
 The shaded areas highlight the much increased overlap between the warp-transformed densities and the reference density $p_2=N(0,1)$. }}
\end{center}\end{figure}

The next obvious transformation is to  match both the location and the spread.
Let $\mu_i$ be a location parameter and $\scale_i$ be a scaling parameter, for $i=1,2$. The Warp-II transformation   then sets
$
\tilde{w}^{\pII}_{i,j}=\scale_i^{-1}(w_{i,j}-\mu_i)
$
and
$
\tilde{q}^{\pII}_i(\omega)=|\scale_i|q_i(\scale_i\omega+\mu_i).
$
%and the corresponding estimator is called the Warp-II bridge sampling.
%\begin{equation*}
%\hat{r}_{\alpha}^{\pII}%=\dfrac{n_2^{-1}\sum_{j=1}^{n_2}\tilde{q}^{\pII}_1(\tilde{w}^{\pII}_{2,j})\alpha(\tilde{w}^{\pII}_{2,j})}{n_1^{-1}\sum_{j=1}^{n_1}\tilde{q}^{\pII}_2(\tilde{w}^{\pII}_{1,j})\alpha(\tilde{w}^{\pII}_{1,j})}
%=\dfrac{n_2^{-1}\sum_{j=1}^{n_2}|\scale_1|q_1(\scale_1\scale_2^{-1}
%(w_{2,j}-\mu_2)+\mu_1)\alpha(\scale_2^{-1}(w_{2,j}-\mu_2))}
%{n_1^{-1}\sum_{j=1}^{n_1}|\scale_2|q_2(\scale_2\scale_1^{-1}(w_{1,j}-\mu_1)+
%\mu_2)\alpha(\scale_1^{-1}(w_{1,j}-\mu_1))}.
%\end{equation*}
The dash-dot curve in the left panel of Figure \ref{warp2} illustrates that  $\tilde{p}_1^\pII$ overlaps more  with $p_2$ than $p_1$ does. It also overlaps more than the Warp-I transformed density $\tilde{p}_1^\pI$ does (not shown).

Warp-III transformations increase overlap further by making the densities in question symmetric via a \textit{stochastic transformation}. Specifically, a Warp-III transformation sets $\tilde{w}^{\pIII}_{i,j}=\xi_j\scale_i^{-1}(w_{i,j}-\mu_i)$, where $\xi_j$ takes on the value $1$ or $-1$ with equal probability (independently of $w_{i,j}$).
%where $b_{i,j}$ takes $-1$ and $1$ with equal probability.
The unnormalized density of $\tilde{w}^{\pIII}$ is
$
\tilde{q}_i^{\pIII}(\omega)=\left|\scale_i\right|\left[q_i\left(\mu_i-\scale_i\omega\right)+q_i\left(\mu_i+\scale_i\omega\right)\right]/2,
$
an example of which is shown in the right panel of
Figure \ref{warp2} (dash-dot curve). %\textcolor{red}{Warp-III in high dimension?}
 Below we show that  stochastic transformations are also very powerful in dealing multi-modality, a challenging issue in MC based estimation and indeed in statistical inference more generally.

%\textcolor{red}{
%In the remainder of this paper, we refer
%the warp-$\mathcal{X}$ bridge sampling to
%the bridge sampling with $w_{i,j}$ and $q_i$ in (\ref{general_bridge_sampling}) replaced
%with the warp-$\mathcal{X}$ transformed samples, $\tilde{w}^{\pX}_{i,j}$, and their corresponding densities, $\tilde{q}_i^\pX$,
%where the superscript $^\pX$ represents the type of transformation.
%The corresponding estimator is denoted as
%$\hat{\lambda}_{\alpha}^{\pX}=\log\left(\hat{r}_{\alpha}^{\pX}\right)$ for general choices of $\alpha$,
%%$\hat{\lambda}_{\geo}^{\pX}=\log\left(\hat{r}_{\geo}^{\ptextX}\right)$ for the geometric bridge sampling
%and $\hat{\lambda}_\opt^{\pX}=\log\left(\hat{r}_\opt^{\pX}\right)$ for
%the optimal bridge sampling.
%%Let $\bz$ be the vector of parameters that characterizes
%%the warp-$\mathcal{X}$ transformation.
%It is important to note that $\alpha$
%is typically a functional of the two densities, as in the geometric
%bridge sampling and the optimal bridge sampling, so
%$\alpha$ may also depend
%on the transformed densities
%in the warp-$\mathcal{X}$ bridge sampling. }

%%%%%%%%%%%%%%%%%%%%%%%%%%%%%%%%%%%%%%%%%%%%%%%%%%%%%%
\section{Warp-U Bridge Sampling}
\label{sec:warpu}

%Recall that the goal of transformation is to increase the overlap of the densities of the transformed data, so that the
%asymptotic variance of $\hat{\lambda}_{\alpha}$ will be reduced while the normalizing constants remain unchanged.

Consider a unimodal density, $\phi$, such as a $\Normal(0,I_d)$ or \textit{t}-distribution. The key idea of our approach is to construct a stochastic transformation of the original MC draws such that the density for the transformed draws is much closer to $\phi$.
To simplify the exposition, we consider the problem of estimating a single  normalizing constant and fix the other density used in the bridge sampling estimator (\ref{general_bridge_sampling}) to be  $\phi$.   The problem of estimating  a  ratio of two normalizing constants  can then be handled in the following two ways. Firstly, we can use two
bridge sampling estimators, one in the numerator and one in the denominator, based  on the Warp-U transformed draws $\{\tilde{w}_{i,1},\ldots, \tilde{w}_{i,n_i}\}\sim \tilde{p}_i$ and draws from the convenient unimodal \textit{auxiliary} distribution $\{z_{i,1},\ldots,z_{i,m_i}\}\sim \phi$, for
$i=1,2$  respectively. The two estimators can share the same auxiliary distribution $\phi$, or even the same set of auxiliary draws.  We emphasize again here that we do not require any of these draws to be i.i.d., though typically those from the auxiliary distribution are i.i.d. because it is  usually easy to obtain i.i.d. draws from $\phi$.  Secondly, we could disregard $\phi$ after the warp transformation and then use one bridge sampling estimator of the ratio $r$  based only on the full set of the transformed draws $\bigcup_{i=1}^2\{\tilde{w}_{i,1},\ldots, \tilde{w}_{i,n_i}\}$. %\sim \tilde{p}_i$, for $i=1,2$.
%which also bypasses the problem of different dimensionalities of the two densities in bridge sampling \citep{chen1997estimating}.
This second strategy is  effective  because
 if $\tilde{p}_1$ and $\tilde{p}_2$ both overlap significantly  with $\phi$ then they are likely to also have substantial overlap with each other. %, see Appendix \ref{sec::estimating_c_1_over_c_2}. %First, the goal of transformation is clear: to transform $w_{i,j}$ so that the density $\tilde{p}_i$ of the transformed data $\tilde{w}_{i,j}$
% will be ``close" to $\phi$, for $j=1, \ldots, n_i$.
%Second, $c_1/c_2$ can also be obtained
%by estimating $c_i$ separately
%with the bridge sampling estimators based on  data from $\tilde{p}_i$ and $\phi$, which also bypasses the problem of
%different dimensionalities of the two densities in bridge sampling \citep{chen1997estimating}.
%Finally, transforming both data sets to have more overlap with a common $\phi$
%often results in substantial overlap between $\tilde{p}_1$ and $\tilde{p}_2$,
%thus $c_1/c_2$ can be well estimated by the bridge sampling with the transformed data.
%

Since we focus on a single unnormalized density  $q$, we drop the double indices and let
$\{w_1, \ldots, w_n\}$
be $n$ draws from $p=c^{-1}q$, where $p$ is assumed to be a continuous density on $\mathbb{R}^d$.
Similarly, we use  $\{z_1, \ldots, z_m\}$ to denote $m$ i.i.d. draws from $\phi$.
For  concreteness, we set $\phi=\Normal(0,I_d)$, but other choices of $\phi$ can work equally well or even better.

%To reduce the asymptotic variance of the bridge sampling estimator, we transform
%$w_j$ so that the density of the transformed data $\tilde{w}_j$, denoted as $\tilde{p}$,
%will be ``close" to $\phi$ in terms of their Harmonic (and possibly other)
%distance.
%

%%%%%%%%%%%%%%%%%%%%%%%%%%%%%%%%%%%%%%%%%%%%%%%%%%%%%%
\subsection{Constructing Warp-U Transformations}\label{intuition_warpU}

When $q$ is multi-modal, we could approximate it by a Gaussian mixture  $\phi\mix$ and then perform standard bridge sampling using $q$ and $\phi\mix$. Warp-U bridge sampling aims to improve on this approach but begins in the same way. %, which we now  also make use of an approximating distribution $\phi\mix$ but improve on transformations improve on this approach, but start in
Let
\begin{equation}\label{equ::mixture_density_general}
\phi_{\mix}(x;\bz)=\sum_{k=1}^K\phi^\pk(x) =\sum_{k=1}^K\pi_k\left|\scale_k\right|^{-1}\phi\left(\scale_k^{-1}(x-\mu_k)\right),
\end{equation}
where $\phi^\pk$ represents the $k$-th component in $\phi_{\mix}$, including its weight $\pi_k$, for $k=1,\ldots, K$, and $\bz$ collects the transformation parameters $\bigcup_{k=1}^K\{\pi_k, \mu_k,\scale_k\}$.
\citet{alspach1972nonlinear} showed that any piecewise continuous density can be approximated arbitrarily well by a Gaussian mixture of the form (\ref{equ::mixture_density_general}) as $K\rightarrow \infty$ (specifically, they demonstrated uniform convergence).
In practice, for a reasonable choice of $K$, it is usually possible to find a $\phi_{\mix}$ that has substantial overlap with $p$.
Section \ref{specific_method} will discuss how to estimate $\phi_\mix$.  Here we
assume that $\phi_\mix$ is known in order to describe  the  Warp-U transformation itself.

\begin{figure}[t]
\begin{center}
\includegraphics[width=0.9\textwidth,trim=28mm 84mm 28mm 72mm,clip]{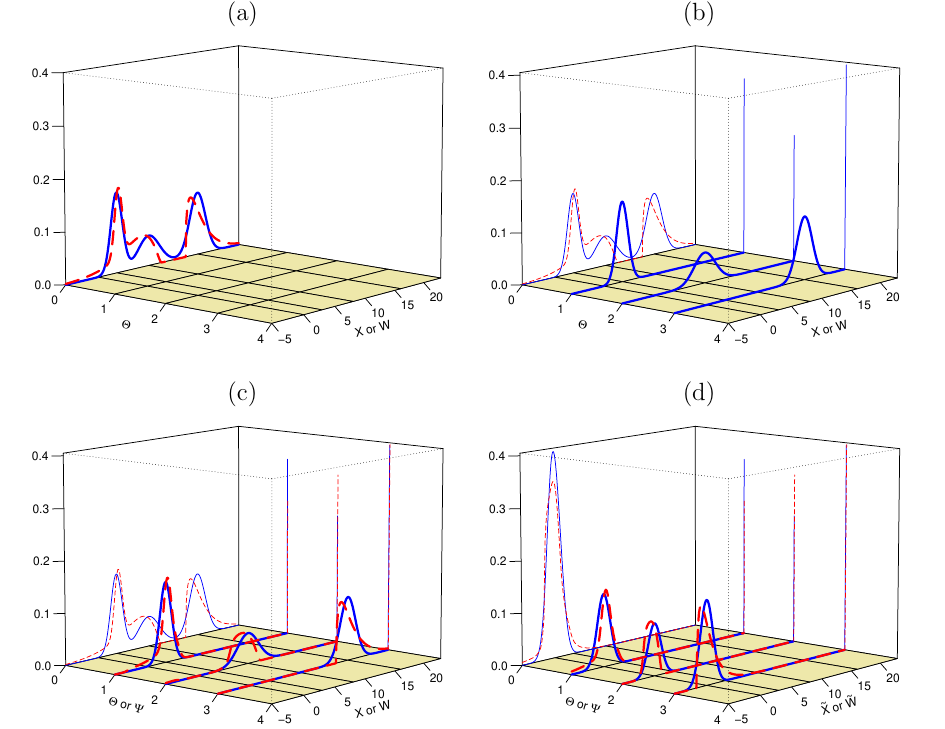}
\caption{\label{animation}\small {Illustration of Warp-U transformation.
(a) $\phi_{\mix}$ (solid line) and $p$ (dashed line); (b) the joint and marginal distributions of
$X$ and $\Theta$ (solid line); (c) the joint and marginal distributions of $W$ and $\Psi$ (dashed line);
(d) the joint and marginal distributions of $\Theta$ and $\widetilde{X}$ (solid line) and those of
$\Psi$ and $\widetilde{W}$ (dashed line), where  $\widetilde{X}$ and $\widetilde{W}$ are obtained via Warp-U transformation.
}}
\end{center}\end{figure}

%From another prospective, if
The Warp-U transformation uses a coupling between augmented random variables drawn from  $\phi\mix$ and $p$, and we now specify this relationship. Suppose $X\sim\phi_{\mix}$, depicted in Figure~\ref{animation}(a) as the solid line, then we can write $X=\scale_{\Theta}Z+\mu_{\Theta},$
where $Z\sim\phi$ and is independent of $\Theta$, a discrete random variable distributed such that
$P(\Theta=k)=\pi_k$ for $k=1, \ldots, K$.
Figure \ref{animation}(b) shows the joint distribution of $\Theta$ and $X$,
with their marginal distributions  on the
two faded vertical plates.
%$P(\theta=i)=\pi_i, \text{ for }i=1,2,3.$
%For $k\in\{1,2,3\}$, we define a deterministic function $\tr_k(x;\bz) = \scale_{k}^{-1}(x-\mu_{k})$.
The random index $\Theta$ induces a
random transformation
 \begin{equation}\label{eq:tran}
\tr_{\Theta}(x) =\scale_{\Theta}^{-1}(x-\mu_{\Theta}).
\end{equation}
It follows trivially that if we draw $(x, \theta)$ from the joint distribution of $(X,\Theta)$,
then $\tilde{x} =\tr_{\theta}(x)\sim  \phi$.
%In other words, the random transformation $f_{\theta}(\cdot;\bz)$ applied to $x_i$ resulted in a random draw
%from $\phi$.
%The joint distribution of $(X, \theta)$ is well defined, so we can create a random variable to be $\tilde{X}=\sigma_{\theta}^{-1}(X-\mu_{\theta})$. Apparently, $\tilde{X}=Z$ follow a standard normal distribution.

%Now we describe how to wrap  $p$ into $\widetilde{p}$, the dashed line in Figure \ref{intuition}.
%L
Next, let $W$ be a random variable from $p$ and $\Psi$ be a random index. We create a coupling between $(W,\Psi)$ and $(X,\Theta)$ by requiring  that $\Psi|W$ and $\Theta|X$ have the same distribution, i.e.,
%\footnote{This is a coupling because, in contrast to the case where the random index $\Psi$ is independent of $W$, our choice of $\Psi$ means that the distribution of $(W,\Psi)$ is similar to that of $(X,\Theta)$, which is key for our method.}.
%That is, we set  %Specifically,  we specify $\Psi$ that the conditional distribution given $W=w$ is the same as the conditional distribution of $\Theta$ given $X=w$.
%create a joint distribution of $W$ and a random variable $\Psi$, so that
%the random transformation $f_{\Psi}(\cdot;\bz)$ can be applied to $W$.
%Since $\phi_{\mix}$ and $p$ have substantial amount of overlap, we can
%specify the conditional distribution of $\Psi$ given
 % the same as the conditional distribution of $\Theta$ given $X$, that is,
\begin{equation}\label{pk}
\varpi(k|\omega) \triangleq  P(\Psi=k|W=\omega) \equiv P(\Theta=k|X=\omega) = {\phi^\pk(\omega)}/{\phi_{\mix}(\omega)},  \quad  k=1,\ldots, K.
\end{equation}
We can then decompose $p$ into $K$ components, i.e., $p(\omega) = \sum_{k=1}^K p^\pk(\omega)$,
where %$p^\pk$ is the joint distribution of $(W,\Psi=k)$, that is,
\begin{equation}\label{equ::p_pk_omega}
p^\pk(\omega) = p(\omega,\Psi=k) = p(\omega)\frac{\phi^\pk(\omega)}{\phi_{\mix}(\omega)}.
\end{equation}
Figure \ref{animation}(c) shows the joint distribution of $(W,\Psi)$
(thick  dashed curves) and their marginal distributions (thin  dash curves
in the two vertical plates). The Warp-U transformation is then constructed by again using the map in (\ref{eq:tran}) but with $(W, \Psi)$ in place of $(X, \Theta)$,
\begin{equation}\label{equ::Warp-U_transformation}
\widetilde{W}=\tr_{\Psi}(W)=\scale_{\Psi}^{-1}(W-\mu_{\Psi})\sim \widetilde{p}.
\end{equation}
Intuitively, because the transformation $\tr$ maps the multi-modal $\phi\mix$ back to the original uni-modal (generating) density $\phi$, when it is applied to the multi-modal $p$, it can achieve similar a  ``uni-modalizing" effect because $\phi\mix$ was chosen to approximate $p$.

In practice, to apply a Warp-U transformation to  $w_j$, we calculate  $\varpi(\cdot|w_j)$
according to (\ref{pk}),
draw $\psi_j$ from $\varpi(\cdot|w_j)$, and finally apply the deterministic transformation $\tr_{\psi_j}$ to $w_{j}$.
Graphically,
each $p^\pk$ in Figure \ref{animation}(c) is re-centered and re-scaled, like
its counterpart, $\phi^\pk$.
The  dashed lines in
Figure \ref{animation}(d) are
the joint distribution of $\Psi$ and the Warp-U transformed variable, $\widetilde{W}$. In the faded left vertical panel of \ref{animation}(d),   we see that the distribution of $\widetilde{W}$ overlaps substantially with $\phi$.

When $K=1$, the Warp-U transformation  is the same as the Warp-II transformation provided that we choose $\phi_{\mix}$ to be a
location-scale family. For $K>1$,
Theorem 1 in Section \ref{theorem} (below) ensures  that there will be additional overlap  between $\tilde{p}$ and
$\phi$  compared to the overlap between  $p$ and $\phi_{\mix}$, except for in trivial cases, e.g., when $p=\phi_{\mix}$. % already or the warp transformation is otherwise a trivial one.
\subsection{Theoretical Guarantee for General Warp-U Transformations}\label{theorem}
\begin{figure}[t]
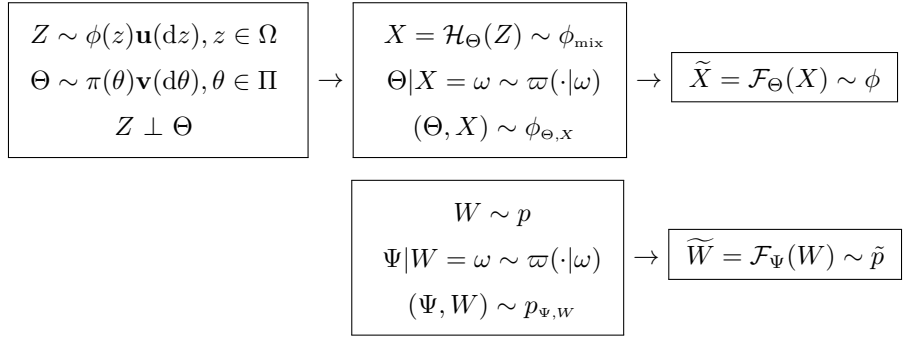

\begin{center}
%\vspace{0.2cm}
\fbox{$
\begin{array}{c}
Z \sim \phi(z)\measure(\ud z), z\in\Omega \\
\Theta\sim\pi(\theta)\measurev(\ud \theta),\theta\in\Pi\ \\
Z~\bot~\Theta
\end{array}$
} $\rightarrow$ \fbox{
$\begin{array}{c}
X=\trh_{\Theta}(Z)\sim \phi_{\mix}\ \\
%\textcolor{red}{\theta|X=\omega \sim \varpi(\cdot|\omega)}
\Theta|X=\omega \sim \varpi(\cdot|\omega)\ \\
(\Theta,X)\sim \phi_{\text{\tiny$\Theta, \hspace{-0.06cm} X$}}\
\end{array}$}
$\rightarrow$
\fbox{
$\widetilde{X}=\tr_{\Theta}(X)\sim \phi$
}\\ \vskip .1in
\hspace{4.6cm}
\fbox{
$\begin{array}{c}
W \sim p\\
%\textcolor{red}{}
\Psi|W=\omega \sim \varpi(\cdot|\omega)\\
(\Psi,W)\sim p_{\text{\tiny$\Psi, \hspace{-0.06cm} W$}}
\end{array}$}
$\rightarrow$
\fbox{
$\widetilde{W}=\tr_{\Psi}(W)\sim \tilde{p}$
}
\caption{\label{representation}\small  {Relationships among the random variables and their distributions   for
Warp-U transformation. Here  for almost surely (with respect to $\measurev$) all values of
$\theta\in\Pi$, $\tr_\theta$ and  its inverse $\trh_\theta$ are  one-to-one, onto, and  almost surely
(with respect to $\measure$) differentiable  maps  from $\Omega\rightarrow \Omega.$   }}
\end{center}\end{figure}

Figure~\ref{representation} summarizes the key variables and distributions underlying a general Warp-U transformation, which does not assume that $\phi$ is the Normal density. We do still require that $\phi$ shares the same support $\Omega$ as our target $p$. Another generalization included in Figure~\ref{representation} is that
 the  ``index variable"
$\Theta$ (and hence also $\Psi$) is permitted to take on any distribution
$\pi$ with support $\Pi$ and dominating measure $\measurev$, and in particular $\Theta$ (and $\Psi$) is no longer required to be discrete.

For all $\theta \in \Pi$, the map $\tr_{\theta}$  in Figure~\ref{representation} is required to be one-to-one, onto, and almost surely
differentiable,  and to satisfy $ \Omega=\tr_\theta(\Omega)$.
We denote its inverse map by  $\trh_{\theta}$. Since we specify $X \sim \trh_{\theta}(Z)$, where $Z\sim \phi$, the
conditional distribution $X\big{|}\Theta=\theta$
 is %then\footnote{Note here $\phi$ is used as a genetic notation, hence it is not necessarily the pdf of $\Normal(0, I_d)$.}
\begin{equation}\label{conditionaldensity}
\phi_{\text{\tiny$X|\Theta$}}(x|\theta) = \phi(\tr_{\theta}(x))\left|\tr_{\theta}'(x)\right|,\quad x\in\Omega
\end{equation}
and the (marginal) density of $X$ is
\begin{equation}\label{mixture_density_expression}
\phi_{\mix}(x) = \int_{\Pi}\phi_{\text{\tiny$X|\Theta$}}(x|\theta)\pi(\theta)\measure(\ud\theta)= \int_{\Pi} \phi(\tr_{\theta}(x))\left|\tr_{\theta}'(x)\right|\pi(\theta)\measure(\ud\theta).
\end{equation}
Let  $\varpi(\cdot|x)$  be the conditional distribution $\Theta|X=x$,
\begin{equation}\label{equ::varpi_theta_omega}
\varpi(\theta|x) = \frac{\phi_{\text{\tiny$X|\Theta$}}(x|\theta)\pi(\theta)}{\phi_{\mix}(x)}, \quad \theta\in\Pi,
\end{equation}
and, as before, let the variable $\Psi$ be defined through %defining the conditional distribution to be
$P(\Psi=\theta|W=\omega)=\varpi(\theta|\omega)$.
%Since $\trh_{\theta}(\cdot)$ is a one-to-one function for any $\theta$, the inverse function, denoted as $f_{\theta}(\cdot)$ or $\tr_{\theta}(\cdot)$, exists.
%Let $\tr_{\theta}(\cdot)$ be the inverse function of $\trh_{\theta}(\cdot)$ for any $\theta$.
The  joint distributions of $(\Psi,W)$ and $(\Theta,X)$ therefore share the same conditional specification:\begin{equation}
\label{equ::joint_dis_psi_w}
p_{\text{\tiny$\Psi, \hspace{-0.06cm} W$}}(\theta,\omega) = \varpi(\theta|\omega)p(\omega)\quad \mbox{and}\quad
\phi_{\text{\tiny$\Theta, \hspace{-0.06cm} X$}}(\theta,\omega) =
\varpi(\theta|\omega)\phi_\mix(\omega), \quad (\omega,\theta)\in\Omega\times\Pi.
\end{equation}
Considering this shared structure, here and in what follows we sometimes use the dummy variables $(\omega,\theta)$ to refer to realizations of both $(W,\Psi)$ and $(X,\Theta)$, and prevent confusion through our notation for the density functions in question.

The key consequence of the coupling (\ref{equ::joint_dis_psi_w}) is that the overlap  between  $\phi$ and the density of the Warp-U transformed $W$:  $\widetilde{W}=\tr_{\Psi}(W)\sim \tilde{p}$, is greater than that  between $\phi_\mix$ and $p$.
%Then, the random transformation indexed by $\Psi$ can be applied to $W$
%The inverse function of $\trh_{\theta}(\cdot)$, denoted as $f_{\theta}(\cdot)$, defines a transformation function
%indexed by a random variable.
%Let $\tilde{p}$ be the density of the transformed random variable $\tilde{W}=f_{\eta}(W)$, where
%the joint distribution of $W$ and $\eta$ is defined by making
%$\eta|W=w$ to be identical to $\theta|X=w$, that is,
%\begin{equation}
%p(\eta|W=w)=p(\theta|X=w) = \frac{p_{X|\theta}(w|\theta)\pi(\theta)}{\int p(w|\theta')\pi(d\theta')}
%\end{equation}
%Let $\varpi(\cdot|x)$ be the conditional distribution of $\theta$ given $X=x$, expressed as
%\begin{equation}
%\varpi(\theta|x) = p(\theta|X=x) = \frac{p(x|\theta)\pi(\theta)}{\int p(x|\theta')\pi(d\theta')}
%\end{equation}
%We define a new random variable $\eta$ by specifying the conditional
%distribution of $\eta$ given $W$ to be
%$p(\eta|W=\omega) =\varpi(\eta|\omega). $
%We apply the random transformation $f_{\eta}(\cdot)$ to $W$ and  $\tilde{W}=f_{\eta}(W)$
To prove this mathematically, we need a measure or multiple measures of overlap.  The notion  of $f$-divergence, or more
precisely its complement (since small divergence corresponds to large  overlap),  serves well  for our purposes. For any (non-trivial)  \textit{convex} function $f$ on $[0, \infty)$ such that
 $f(1)=0$, the corresponding $f$-divergence between two probability densities  $p_1$ and $p_2$, when
 $p_1$ is absolutely continuous with respect to $p_2$,  is defined as
\begin{equation}\label{eq:fdiv}
\mathcal{D}_f(p_1||p_2) = \int_\Omega p_2(\omega)f\left(\frac{p_1(\omega)}{p_2(\omega)}\right)\measure(\ud\omega).
\end{equation}
Theorem 1 below states that Warp-U transformations reduce  any $f$-divergence, unless either the transformation or $f$  is  trivially chosen, in which cases the $f$-divergence is unchanged. The proof is given in Appendix \ref{app:proof}. \vspace{0.2cm}

{\bf Theorem 1.} {Let the Warp-U transformation  $\tr_{\Psi}$ be defined as in Figure~\ref{representation}, with the conditions given in the caption. The following results then hold.}
\begin{itemize}
\item[(I)] For any $f$-divergence $\mathcal{D}_f$, we have  \
$%\label{equ::main_inequality}
\mathcal{D}_f(\tilde{p}||\phi)\leqslant \mathcal{D}_f(p||\phi_{\mix}).
$
\item[(II)] If $f$ is strictly convex,  then   the equality in (I) holds if and only if  $\ell(\theta;\tilde \omega)\equiv\frac{p(\trh_{\theta}(\tilde\omega))}{\phi_{\mix}(\trh_{\theta}(\tilde \omega))}$ is free of $\theta$ (almost surely with respect to $\measurev\times\measure$).
\end{itemize}

The Hellinger distance, the weighted harmonic divergence in (\ref{harmonic_distance}), and the $L_1$ distance are all $f$-divergences, with $f_{He}(t)=0.5(1-\sqrt{t})^2$, $f_{Ha}(t)=w_1(1-t)/(w_1+w_2t)$, and $f_{L_1}(t)=|1-t|$, respectively. The weighted harmonic divergence in (\ref{harmonic_distance}) is an especially important case because it determines the asymptotic variance of bridge sampling estimators (see (\ref{opt_variance})). Consequently, Theorem 1 says that the  bridge sampling estimator based on $\tilde{p}$ and $\phi$ has smaller asymptotic variance than that based on $p$ and $\phi_{\mix}$, thus supporting the use of Warp-U transformations (Section \ref{subsection::simulation_known_mixture} below gives the explicit form of these two estimators).
Interestingly, inequality (I)
does not necessarily hold for $L_{\text{p}}$ distance when $\text{p}\neq1$ (and hence $L_{\text p}$ distance is not an $f$-divergence when ${\text p}\not=1$). As a simple counter-example, let $K=1$ in (\ref{equ::mixture_density_general})
and therefore  $\phi_{\mix}(\omega)=|\scale|^{-1}\phi\left(\scale^{-1}(\omega-\mu)\right)$.  Then $\tilde{p}(\omega)=|\scale| p(\scale \omega+\mu)$, and
\begin{equation*}
\begin{split}
L_{\text{p}}(\tilde{p}, \phi)&=
\left(\int \left||\scale| p(\scale \tilde{\omega}+\mu)-\phi(\tilde{\omega})\right|^{\text{p}}\measure(\ud\tilde{\omega})\right)^{\text{p}^{-1}}
=|\scale|^{1-\text{p}^{-1}}L_{\text{p}}(p,\phi_{\mix}),
\end{split}
\end{equation*}
so $L_{\text{p}}(\tilde{p},\phi)> L_{\text{p}}(p,\phi_{\mix})$ whenever  $|\scale|^{1-{\text p}^{-1}}>1$ (and $ L_{\text{p}}(p,\phi_{\mix})>0$).

Part (II) of Theorem 1 means that a Warp-U transformation will always result in real gain, as measured by any strictly convex $f$-divergence,  unless one of two situations occur: (A)  $\phi_{\mix}$ is a perfect fit to $p$, in which case obviously $\ell(\theta,\tilde \omega)=1$; or (B) $p\not= \phi_{\mix}$, but the Warp-U transformation  $\tr_\Theta$ is  unfortunately  (or unwisely) chosen such that   it renders the ``likelihood ratio" $\ell(\theta;\tilde\omega)$ flat as a function of $\theta$.  Situation (B) includes the trivial  cases where $\tr_\theta$ does not depend on $\theta$, or $\theta$ does not vary because  $\pi$ is a singleton, as well as some more complex scenarios.

\begin{figure}[t]
\begin{center}
\includegraphics[width=1\textwidth,trim=40mm 92mm 40mm 220mm,clip]{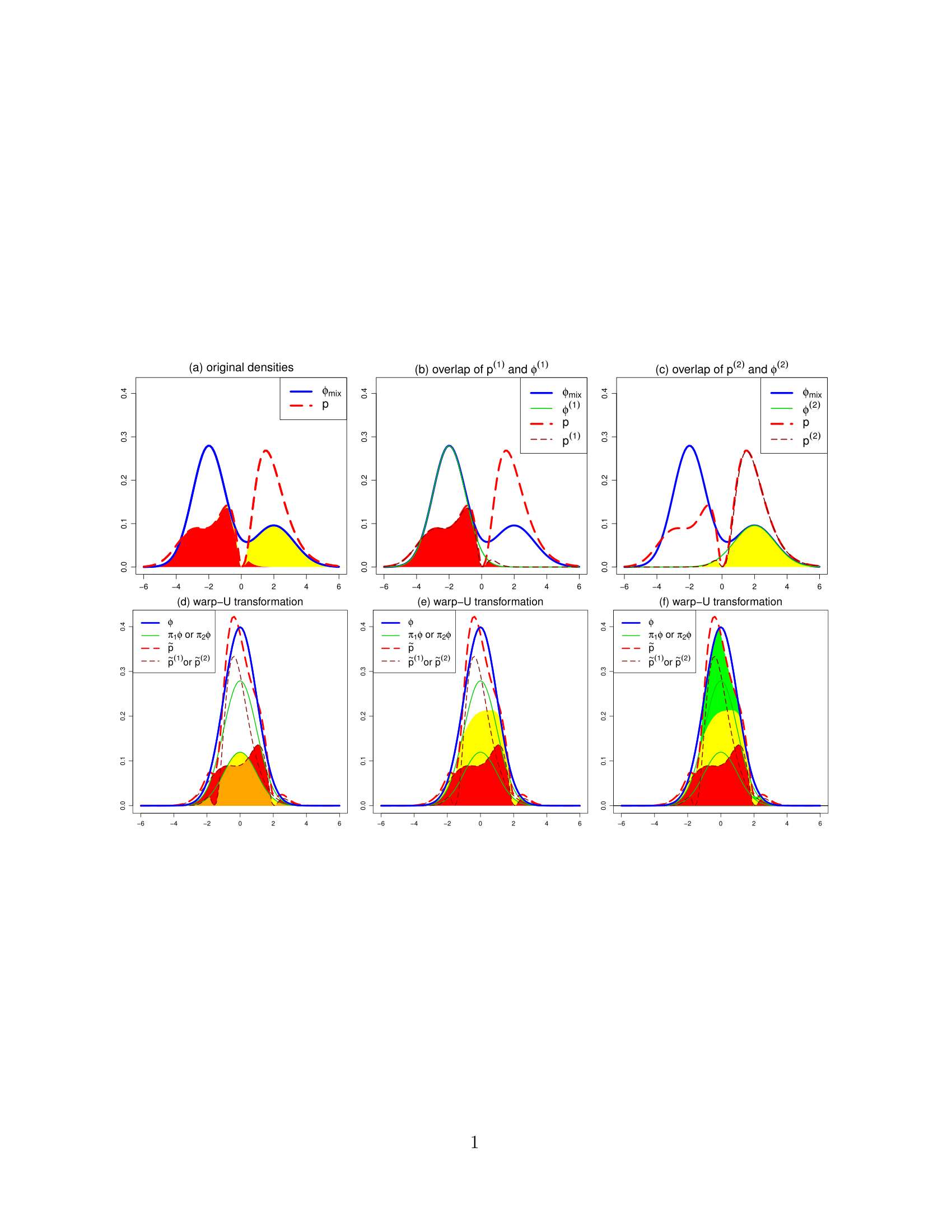}
%\caption{Graphical illustration of the increase of the overlap after Warp-U transformation.}
\vskip -1.8in
\caption{\label{graphical_illustration}\small{ Illustration of  the increase in the area of the overlapping region
after Warp-U transformation.
(a) $p$ (dashed line)  and $\phi_{\mix}$ (solid line); (b)
the 1st component of $p$,  denoted as $p^{\text{\tiny$(1)$}}$ (thin dashed line),
the 1st component of $\phi_{\mix}$, denoted as $\phi^{\text{\tiny$(1)$}}$ (thin solid line),
and their overlap (shaded in red);
(c) $p^{\text{\tiny$(2)$}}$,  $\phi^{\text{\tiny$(2)$}}$, and
their overlap (shaded in yellow); (d) the corresponding curves and shaded areas after Warp-U transformation;
(e) the yellow region is added on top of the red region; (f) the green area shows the additional
cross overlap between the 1st and 2nd components induced by the Warp-U transformation.
}}
\end{center}\end{figure}

An illustration of Theorem 1 is given in Figure \ref{graphical_illustration} and Table \ref{distancetable} for the case of a tri-modal target distribution $p$ and an approximating density $\phi_\mix$ with $K=2$. The green region in Figure \ref{graphical_illustration}(f) shows that the final overlap between $\tilde{p}$ and $\phi$ after Warp-U transformation is greater than the sum of the overlaps between $p^{(k)}$ and $\phi^{(k)}$, for $k=1,2$. We call the additional overlap {\it cross-overlap}, and it is this phenomenon which is key to Warp-U transformations. Table \ref{distancetable} lists several $f$-divergences for the pairs $(p,\phi_{\mix})$ and $(\tilde{p},\phi)$,  and it confirms that the $f$-divergences are much lower in the latter case. Indeed, due to cross-overlap, the overlapping area for the pair of densities $(\tilde{p},\phi)$ is nearly  40\% larger than that for $(p,\phi_{\mix})$.

\begin{table}[h]
  \caption{\label{distancetable}\small  {The overlapping area, and the distances between
$p$ and $\phi_{\mix}$ and  between $\tilde{p}$ and $\phi$.}}
\vspace{-0.6cm}
\begin{center}
\resizebox{1\textwidth}{!}{
  \begin{tabular}{|l|c|c|c|c|}
\hline
Densities   & Overlap area & $~L_\text{1}$ distance~ & Hellinger distance  & harmonic divergence \\\hline
 $(p,\phi_{\mix})$  & 0.66 & 0.68   & 0.28  & 0.145\\\hline
$(\tilde{p},\phi)$   & 0.92  &  0.16  & 0.08  & 0.013\\
\hline
\end{tabular}}
\end{center}
\end{table}

\subsection{Warp-U Bridge Sampling}\label{subsection::simulation_known_mixture}

After the parameters $\bz$ for $\phi_\mix$ have been chosen, the Warp-U transformation is determined. The unnormalized density of the transformed draws $\{\tilde{w}_1,\ldots,\tilde{w}_n\}$ can then be expressed as
\begin{equation}\label{equ::Warp-U-transformed-density}
\tilde{q}(\tilde{w};\bz)%=\sum_{k=1}^K\varpi(k|\scale_k\omega+\mu_k)q(\scale_k\omega+\mu_k)|\scale_k|
=\sum_{k=1}^K cp^{(k)}(\omega=\scale_k\tilde{w}+\mu_k)%=\sum_{k=1}^K cp^{(k)}(\scale_k\tilde{w}+\mu_k)
%=\sum_{k=1}^K c\tilde{p}^{(k)}(\tilde{w})
= \phi(\tilde{w})\sum_{k=1}^K\frac{q(\scale_k\tilde{w}+\mu_k)}{\phi_{\mix}(\scale_k\tilde{w}+\mu_k)}\pi_k.
\end{equation}
%where $\bz$ denotes the vector of parameters in $\phi_\mix$.
%By replacing $q$ with $cp$, we get $\tilde{q}= c\tilde{p}$, meaning
Clearly, the normalizing constants of $\tilde{q}$ and $q$ are both $c$, and hence
%\begin{equation}\begin{split}
%\int \tilde{q}(\omega;\bz)d\omega&=
%\sum_{k=1}^K\int \phi(\omega)\frac{q(\scale_k\omega+\mu_k)}{\phi_{\mix}(\scale_k\omega+\mu_k)}\pi_k\ud\omega\\&= \sum_{k=1}^K\int \pi_k|\scale_k|^{-1}\phi\left(\scale_k^{-1}(\tilde{\omega}-\mu_k)\right)\frac{q(\tilde{\omega})}{\phi_{\mix}(\tilde{\omega})}\ud\tilde{\omega}\\
%&=\int \left[\sum_{k=1}^K \pi_k|\scale_k|^{-1}\phi\left(\Sigma_k^{-1}(\tilde{\omega}-\mu_k)\right)\right]\frac{q(\tilde{\omega})}{\phi_{\mix}(\tilde{\omega})}\ud\tilde{\omega}
%%\sum_{k=1}^K\int \varpi(k|\Sigma_k\omega+\mu_k)q(\Sigma_k\omega+\mu_k)\Sigma_kd\omega=\sum_{k=1}^K\int \varpi(k|\omega)q(\omega)d\omega\\
%%&=\int \left(\sum_{k=1}^K\varpi(k|\omega)\right)q(\omega)d\omega%\stackrel{\mathrm{*}}{=}
%=\int q(\omega)d\omega,
%\end{split}\end{equation}
%where the second to last equality is obtained by plugging the expression of $\phi_{\mix}$ in
% (\ref{equ::mixture_density_general}).
we can estimate $c$ with the bridge sampling estimator based on
%$\{z_j\}_{j=1}^{m}\stackrel{\mathrm{iid}}{\sim}\phi$ and $\{\tilde{w}_{j}\}_{j=1}^{n}\sim\tilde{p}$ to estimate $c$.
$\{\tilde{w}_1,\ldots, \tilde{w}_n\}\sim\tilde{p}$ and $\{z_1,\ldots,z_m\}\sim\phi$, i.e.,
\begin{equation}\label{equ::warp_u_bridge_sampling}
\hat{c}_{\alpha}^{\pU}\equiv\hat{r}_{\alpha}^{\pU}=\dfrac{m^{-1}\sum_{j=1}^{m}\tilde{q}(z_j;\bz)\alpha(z_j;\tilde{p},\phi)}
{n^{-1}\sum_{j=1}^{n}\phi(\tilde{w}_{j})\alpha(\tilde{w}_{j};\tilde{p},\phi)}.
\end{equation}
As mentioned in Section \ref{sec:bridge_sampling}, the optimal choice of
$\alpha(\cdot;\tilde{p},\phi)$ is proportional to $(s_1\tilde{p}+s_2\phi)^{-1}$.
Since $\phi_{\mix}$ also has some overlap with $p$,
the normalizing constant can alternatively be estimated with the
bridge sampling estimator based on $\{w_1,\ldots,w_n\}\sim p$ and $\{x_1,\ldots,x_m\}\sim\phi_{\mix}$, i.e.,
\begin{equation}\label{equ::mixture_bridge_sampling}
\hat{c}^{\pmix}_{\alpha}\equiv \hat{r}_{\alpha}^{\pmix}=\dfrac{m^{-1}\sum_{j=1}^{m}q(x_j)\alpha(x_j;p,\phi_{{\mix}})}
{n^{-1}\sum_{j=1}^{n}\phi_{{\mix}}(w_{j};\bz)\alpha(w_{j};p,\phi_{{\mix}})}.
\end{equation}
%After the Warp-U transformation specified by $\bz$, $\phi$ and $\tilde{p}$ should have sufficient overlap, as illustrated in Figure \ref{intuition} (right).
%, more than that between $\phi_{\mix}$ and $p$, as the theorem 1 below says.
%How does the relative variance of $\hat{c}$ in (\ref{equ::warp_u_bridge_sampling})
%compared with that of $\hat{c}^*$ in (\ref{equ::mixture_bridge_sampling})?
%Theorem 1 below implies that
%the Harmonic distance and the Hellinger distance between the two densities
%decrease after Warp-U transformation,
%Warp-U transformation guarantees the Harmonic distance between
%$\tilde{p}$ and $\phi$ is smaller than that between $p$ and $\phi_{\mix}$, indicating that
Theorem 1 implies $\mathcal{D}(\tilde{p},\phi)\leqslant \mathcal{D}(p,\phi_{\mix})$ when $\mathcal{D}$ is the weighted harmonic divergence in (\ref{harmonic_distance}),
so the asymptotic variance of $\hat{\lambda}^{\pU}_{\alpha}=\log\left(\hat{c}^{\pU}_{\alpha}\right)$
is smaller than that of $\hat{\lambda}^{\pmix}_{\alpha}=\log\left(\hat{c}^{\pmix}_{\alpha}\right)$
under the optimal choice of $\alpha$, when the draws are independent. Even when we choose some other $\alpha$ (e.g., the geometric mean $\sqrt{p_1p_2}$) or the draws are not independent, we can still expect that the increased overlap obtained by the Warp-U transformation helps (\ref{equ::warp_u_bridge_sampling}) to outperform
(\ref{equ::mixture_bridge_sampling}), at least when both use the same $n$ and $m$. % and we demonstrate this in the next section.

In challenging situations, our choice of $\phi_\mix$ may be a poor match with $p$. This will clearly impact the quality of the corresponding Warp-U  transformation, %and and the Warp-U transformation may move the the components $p^\pk$ farther apart. However,
but $\mathcal{D}(\tilde{p},\phi)\leqslant \mathcal{D}(p,\phi_{\mix})$ will nevertheless still hold. Perhaps the most important case where the Warp-U transformation is not beneficial is when $p$ has many highly isolated modes. The cross-overlap discussed in Section \ref{theorem} will then be small, and so will the reduction achieved by $\mathcal{D}(\tilde{p},\phi)$.   Consequently, in this scenario, the extra computation
required by the Warp-U transformation is not worthwhile. %a standard bridge sampling estimator using $\phi_\mix$ and $p$ will perform almost as well as the Warp-U bridge sampling estimator, see Section \ref{subsection::simulation_known_mixture} below.

\section{Warp-U Computation Details and Numerical Examples}\label{specific_method}

The key step in applying Warp-U bridge sampling is to identify a mixture density $\phi_{\mix}$ that adequately overlaps with $p$, under reasonable constraints on computation.
In relatively low-dimensional ($\leqslant 10$) problems, we can obtain $\phi_{\mix}$ based on the expression for $q$, e.g., using iterated Laplace approximations
\citep[see][]{bornkamp2011approximating,gelman2013bayesian}. However,
these methods are too costly and unstable in  high dimensions.
Below we outline a simple method which uses the draws $\{w_1,\dots,w_n\}$, can capture a good proportion of the mass of $p$,
and has computational cost that is linear in dimensionality. We then adopt another practical strategy to remove an over-fitting bias due to this simple method. %and to approximate the variance of the resulting estimator. %Practical guidance for setting tuning parameters, including $K$, is provided in Appendix \ref{app:tuning}.

%%%%%%%%%%%%%%%%%%%%%%%%%%%%%%%%%%%%%%%%%%%%%%%%%%%%%%
\subsection{Fitting $\phi_{\mix}$: Diagonal Covariance Matrices}\label{sec::EM_algorithm}
%As discussed in the previous section,
%%In the first step of Warp-U transformation,
%it is not necessary (and computationally too expensive) to find a $\phi_{\mix}$ that is almost identical to
%$p$ in applying Warp-U transformation; we only need a $\phi_{\mix}$ that has sufficient overlap with $p$. Then the Warp-U
%transformation will increase the overlap even more.

Suppose that $p$ is $D$ dimensional and that our draws from $p$ reasonably represent the regions of non-negligible density.
We seek a Normal mixture $\phi\mix$ in the form of (\ref{equ::mixture_density_general}) to approximate $p$, where $\scale_k$ is a positive definite \textit{diagonal} matrix, $
\scale_k =\text{Diag}\{\sigma_{k,1}, \sigma_{k,2},  \ldots,  \sigma_{k,D}\}$, for $k=1,\dots,K$,
and hence
$\bz = \left(\pi_1,\ldots,\pi_K,\mu_1,\ldots,\right.$ $\left.\mu_K,
\scale_1,\ldots,\scale_K\right).$ Unlike usual statistical inference problems where ignoring correlations can have very serious consequences,
for Warp-U transformations using diagonal  covariance  matrices is  often an acceptable compromise between
computational efficiency  and MC efficiency.  Indeed, as discussed in Section \ref{sec:warpu}, it is not necessary for
$\phi_{\mix}$ to be a great fit to $p$ in order for us to benefit significantly from Warp-U transformations. %Indeed,  applying Warp-U with reasonable but not perfect $\phi_\mix$ is usually more efficient than increasing $K$ until $\phi_\mix$ is very precise and then applying Warp-U bridge sampling (or standard bridge sampling between $p$ and $\phi_\mix$).
In the next section, we provide further empirical evidence  to illustrate this point.

Since a mixture of Normal components without suitable restrictions
has unbounded likelihood \citep{kiefer1956consistency,day1969estimating},  we estimate $\bz$ by the penalized MLE proposed by \citet{chen2008inference}. In particular,
we make use of the EM procedure proposed by \citet{chen2009inference},  but with a ``robustified" penalty function
\begin{equation*}
\textbf{p}_\textbf{n}(\bz) = -\frac{1}{\sqrt{n}}\sum_{k=1}^K\sum_{d=1}^D \left\{\frac{\hspace{0pt}\widehat{\hspace{0pt}IQ\hspace{0pt}}_{\hspace{0pt}d}^{\hspace{0pt}2}}{\sigma_{k,d}^2}
-\log(\sigma_{k,d}^2)\right\},
\end{equation*}
where $\hspace{0pt}\widehat{\hspace{0pt}IQ\hspace{0pt}}_{\hspace{0pt}d}$
is the inter-quantile range of the  draws from $p$ in the $d$-th dimension. Because EM tends to become trapped  at local modes, we apply it  $M$ times, randomly generating a new initial point  $\bz^{\text{\tiny(0)}}$ for each repetition as follows. The initial values for the $\pi_k$'s and $\scale_k$'s are $\pi_k^{(k)}=K^{-1}$ and $\sigma_{k,d}^2=1.5\widehat{IQ}_d^2$
for all $k$ and all $M$ replications. For the mean parameters $\mu_k$, for the first $M/2$ replications, we randomly sample $K$ of available draws from $p$ (without replacement)  to be the initial values. For the second $M/2$ replications,
along the dimension with the  largest estimated variance, we first identify a region where 95\% of the draws from $p$ reside and divide it into $K$ subregions
so that each subregion contains approximately the same number of draws.  We then sample one draw from
each of the $K$ subregions to set the initial mean parameters.
Our EM stopping criterion is   $|1- (l_n^{(t)}/l_n^{(t-1)})|<10^{-6}$,
where $l_n^{(t)}$ is the value of the (un-penalized) log-likelihood at iteration $t$.
%In all the simulation studies we conducted,
%in 95\% of the times,
In our simulations, the
 EM usually stopped within 100 iterations.
%The weights in $\bz^{\text{\tiny(0)}}$ are all set to be $1/K$, and
%the variance parameters are set to be comparable with the variance of the data.
%Each time, the mean vectors in $\bz^{\text{\tiny(0)}}$ is obtained by
%randomly selecting $K$ data points from $\{w_1,\ldots,w_n\}$.
After obtaining $M$ estimates of $\bz$, we choose the one with the largest likelihood
value to be the parameter, $\widetilde{\bz}$, for Warp-U bridge sampling. Simulations show
that $M$ as small as 2 to 10 is sufficient  to obtain a local maxima that serves well for the purpose of ensuring adequate overlap between $p$ and $\phi_{\mix}$.

\subsection{Overcoming Adaptive Bias and Setting Tuning Parameters}\label{adaptive_bias}

Let $\widetilde{\bz}_{\text{\tiny $\mathcal{D}$}}$
be the estimate of  $\bz$ obtained by applying EM to all the draws from $p$,
$\mathcal{D}=\{w_{1},\ldots,w_n\}$,
and let $\hat{\lambda}^{\pU}_{\text{\tiny $\mathcal{D}$}} = \log\left(\hat{c}^{\pU}_{\text{\tiny $\mathcal{D}$}}\right)$ be the corresponding Warp-U bridge sampling estimator. Because $\widetilde{\bz}_{\text{\tiny $\mathcal{D}$}}$ is a function of the draws from $p$,
the distribution of the corresponding Warp-U transformed draws,
$\{\tilde{w}_{1},\ldots,\tilde{w}_n\}$, is
no longer proportional to  $\tilde{q}(\cdot;\bz)$ in (\ref{equ::Warp-U-transformed-density}) when we substitute $\bz=\widetilde{\bz}_{\text{\tiny $\mathcal{D}$}}.$ In other words, $\hat{\lambda}_{\text{\tiny $\mathcal{D}$}}^\pU$  has an adaptive bias  induced by the dependence of $\widetilde{\bz}_{\text{\tiny $\mathcal{D}$}}$ on $\mathcal{D}$, demonstrated in  Figure~\ref{bias_sd_trade_off} (see Section \ref{sec:examples}).

%\subsection{Overcoming Adaptive Bias Without Increasing Variance}
Since the additional bias of $\hat{\lambda}_{\text{\tiny $\mathcal{D}$}}^\pU$ is due to the dependence of
$\widetilde{\bz}_{\text{\tiny $\mathcal{D}$}}$ on the draws from $p$,
an obvious remedy is to use two disjoint subsets of the draws from $p$
for  estimating $\bz$ and for bridge sampling.
We can then switch the roles of these subsets to gain more statistical efficiency. Figure~\ref{fig::solution_small_bias_variance}
 depicts the sub-sampling strategy we use to obtain
two separate bridge sampling estimators,   $\hat{\lambda}_\text{\tiny H$_i$}^{\pU}, i=1,2$. Each
$\hat{\lambda}_\text{\tiny H$_i$}^{\pU}$ is obtained by using
$L\leqslant n/2$ of the draws  from $p$ to estimate $\bz$ and the other $50\%$ of the draws  for the Warp-U bridge sampling specified by the estimated $\bz$.  Our final estimator $\hat{\lambda}_\text{\tiny H}^{\pU}$ is the
%weighted
average of $\hat{\lambda}_\text{\tiny H$_1$}^{\pU}$ and $\hat{\lambda}_\text{\tiny H$_2$}^{\pU}$. Our empirical investigations, under the setting of i.i.d. draws from $p$ and $\phi$, suggest that the correlation between $\hat{\lambda}_\text{\tiny H$_1$}^{\pU}$ and
$\hat{\lambda}_\text{\tiny H$_2$}^{\pU}$ is often very small
(e.g., $<0.06$). %; see Figure \ref{fig::small_correlation} in Appendix \ref{appendix::property_of_lambda}).
Thus, when the i.i.d. assumption approximately holds, the variance of $\hat{\lambda}_\text{\tiny H}^{\pU}$ is nearly half that of  $\hat{\lambda}_\text{\tiny H$_i$}^{\pU}$.

\begin{figure}
\begin{center}
\begin{tabular}{|c|c|c|c|cccc}
\cline{1-4}
\cellcolor{orange!50}EM & ~~~~~~~~  & \multicolumn{2}{ |c| }{\cellcolor{blue!25} BS} & $\rightarrow$ & $\hat{\lambda}_{\text{\tiny H$_2$}}^{\pU}$ & \multirow{2}{*}{$\rightarrow$} &
\multirow{2}{*}{$\hat{\lambda}_{\text{\tiny H}}^{\pU}=\frac{1}{2}\left(\hat{\lambda}_{\text{\tiny H$_1$}}^{\pU}+\hat{\lambda}_{\text{\tiny H$_2$}}^{\pU}\right)$}  \\
\cline{1-4}
 \multicolumn{2}{ |c| }{\cellcolor{blue!25} BS} & ~~~~~~~~  & \cellcolor{orange!50}EM  & $\rightarrow$ & $\hat{\lambda}_{\text{\tiny H$_2$}}^{\pU}$\\\cline{1-4}
\end{tabular}
\caption{\label{fig::solution_small_bias_variance}\small{A strategy  for  removing the adaptive bias without (unduly) increasing the variance of the Warp-U bridge sampling estimator.
Each  $\hat{\lambda}_{\text{\tiny H$_i$}}^{\pU}, i=1,2$ uses  up to $50\%$ of the draws from $p$ for estimating ${\bz}$ and the other $50\%$ for Warp-U bridge sampling. We then average the two estimators. }}
\end{center}
\end{figure}

As depicted in Figure \ref{fig::solution_small_bias_variance}, we may choose $L<n/2$ in order to reduce the EM computation, which is a reasonable strategy given that $\phi_\mix$ does not need to be a very precise approximation to $p$. Appendix \ref{app:tuning} provides further practical guidance for setting $K$, $L$,  and $m$ (the number of draws from $\phi$), which we now summarize. Firstly, we suggest setting $K\leq n/100$ because our simulations suggest that $\phi_{\mix}$ tends to overfit for $K>n/100$  which can even cause the RMSE of $\hat{\lambda}_{\text{\tiny H}}^\pU$ to increase. Another reason to avoid large $K$ is that computational cost increases quadratically with $K$. Next,  we found that a reasonable choice of $L$ is  $\min(50K,n/2)$ because the reductions in the RMSE of $\hat{\lambda}_{\text{\tiny H}}^\pU$ are relatively small when we increase $L$ past $50K$. Lastly,  in terms of precision per CPU second, when $K$ is already large, increasing $m$ is a more efficient strategy for reducing the variance of $\hat{\lambda}_{\text{\tiny H}}^\pU$  than increasing $K$ further. However, when $K$ is small it is often more efficient to increase $K$ rather than $m$, because if there are fewer components in $\phi_{\mix}$ than there are major modes of $p$, or if the modes of $p$ are asymmetric or heavy tailed, then large reductions in RMSE can usually be obtained by increasing $K$.

%In practice, it is often useful to have a sense of the variance of $\hat{\lambda}_\text{\tiny H}^{\pU}$, and  approximations  can be obtained in multiple ways, including by MC methods. %In Appendix \ref{appendix::property_of_lambda}, we provide a quick assessment under the assumption that all samples are i.i.d. %In future work we would like to verify that the approximation is valid without the i.i.d. assumption.

\subsection{Examples in 10 and 50 dimensions}\label{sec:examples}

To illustrate the effectiveness of using diagonal covariance matrices and the above bias reduction strategy, we first consider a 10 dimensional example where $p$ is set to be a mixture of 25  multivariate skew-{\it t} distributions, whose density is given in the R package `sn' by \cite{azzalini2011r}, also see \citet{azzalini2013skew}.
We specify the degrees of freedom of the 25 skew-{\it t} distributions to take various values between 1 and 4, the skew parameters to take values between $-100$ and  200,
and the scale matrices to be non-sparse. To evaluate our methods properly, we simulate $10^4$ replicate data sets, each of which contains 2500 independent draws from $p$.

We consider three Warp-U bridge sampling estimators, with diagonal covariance matrices for $\phi_{\mix}$. They are $\hat{\lambda}_{\text{\tiny $\mathcal{D}$,$\mathcal{D}$iag}}^{\pU}$, $\hat{\lambda}_{\text{\tiny H,$\mathcal{D}$iag}}^{\pU}$, and
$\hat{\lambda}_{\text{\tiny I,$\mathcal{D}$iag}}^{\pU}$,   where the first subscript
specifies whether $\hat{\lambda}$ is computed by estimating $\bz$ using all the draws from $p$ ($\mathcal{D}$), by setting $L=n/2$ and using the scheme in Section \ref{adaptive_bias} ($\text{H}$), or by estimating $\bz$ from an independent set of draws ($\mathcal{I}$). For all three estimators we use the optimal choice of $\alpha$ and set $m=2500$, i.e., the number of independent draws from the auxiliary  $\phi=\Normal(0,I_{10})$. %\xr{Do the three estimators share the same set of auxiliary draws?\\
%DJ: I don't know the answer to this. The Dropbox does not have the full code or output for these examples. }.
The estimator $\hat{\lambda}_{\text{\tiny $\mathcal{I}$,Diag}}^{\pU}$ serves as a benchmark for comparison because it is free of adaptive bias.

\begin{figure}[t]
\hspace{1.3cm}\begin{minipage}{0.01\textwidth}
\vspace{0.3cm}\rotatebox{90}{\footnotesize Diagonal Cov Matrices}\\\vspace{0.5cm}\\\rotatebox{90}{\footnotesize Full Cov Matrices}\\\vspace{0.1cm}
\end{minipage}\hspace{-1.2cm}\begin{minipage}{0.99\textwidth}
 \begin{center}
\includegraphics[width=0.8\textwidth,trim=32mm 90mm 27mm 83mm,clip]{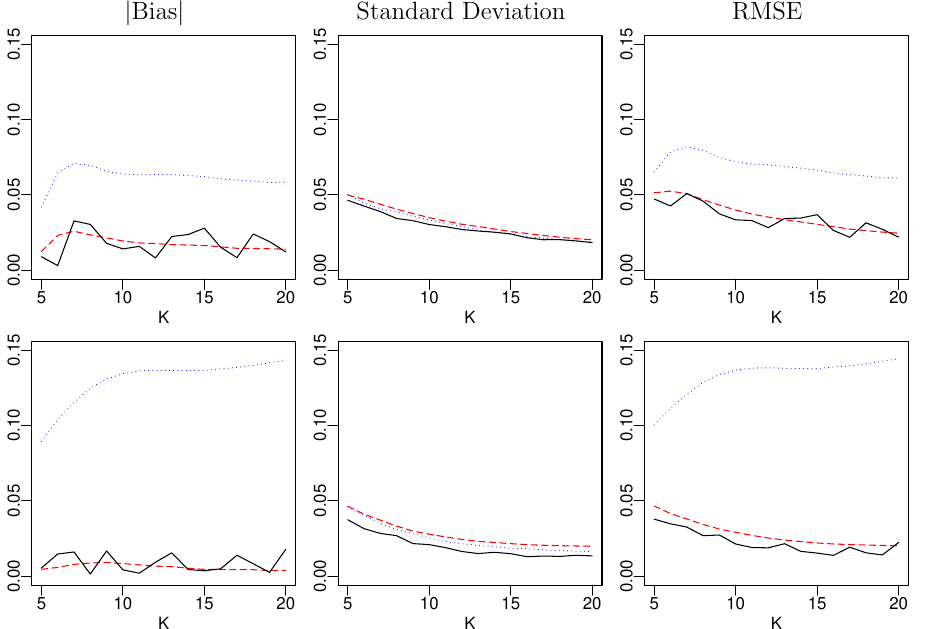}
\end{center}
\end{minipage}
  \caption{\label{bias_sd_trade_off}\small{The columns show the $|$bias$|$,
the standard deviation, and the RMSE of (i)
$\hat{\lambda}_{\text{\tiny $\mathcal{D}$,$\mathcal{Z}$}}^{\pU}=
  \log(\hat{c}_{\text{\tiny $\mathcal{D}$,$\mathcal{Z}$}}^{\pU})$ (dotted lines), the Warp-U  estimator specified by $\widetilde{\bz}_{\text{\tiny $\mathcal{D}$}}$, which is
 estimated from $\mathcal{D}=\{w_1,\ldots,w_n\}$, (ii)
$\hat{\lambda}_{\text{\tiny $\mathcal{I}$,$\mathcal{Z}$}}^{\pU}=
  \log(\hat{c}_{\text{\tiny $\mathcal{I}$,$\mathcal{Z}$}}^{\pU})$   (solid lines), Warp-U  specified by $\widetilde{\bz}_\text{\tiny $\mathcal{I}$}$, which is independent of $\mathcal{D}$, and (iii)
(dashed lines) the average of two Warp-U bridge sampling estimators with
half of the draws from $p$ for estimating $\bz$ and the other half for bridge sampling.
The subscript ``$\mathcal{Z}$" indicates  ``Diag" (top row) or ``Full" (bottom row)
covariance matrices in the
Gaussian mixture model. }}
\end{figure}

The lines in the top row of Figure  \ref{bias_sd_trade_off} show the bias (left panel), the standard deviation (center panel), and the RMSE (right panel) of the three estimators: $\hat{\lambda}_{\text{\tiny $\mathcal{D}$,$\mathcal{D}$iag}}^{\pU}$ (dotted lines), $\hat{\lambda}_{\text{\tiny H,$\mathcal{D}$iag}}^{\pU}$ (dashed lines), and
$\hat{\lambda}_{\text{\tiny I,$\mathcal{D}$iag}}^{\pU}$ (solid lines). Results are plotted for all values of $K$ between $5$ and $20$ inclusive. Larger values of $K$ generally represent a better approximation to $p$ but more computation (and potentially less gain from using Warp-U bridge sampling as opposed to standard bridge sampling between $p$ and $\phi_\mix$). The top left panel of Figure \ref{bias_sd_trade_off} shows the excessive bias
of $\hat{\lambda}_{\text{\tiny $\mathcal{D}$,$\mathcal{D}$iag}}^{\pU}$ compared with $\hat{\lambda}_{\text{\tiny $\mathcal{I}$,$\mathcal{D}$iag}}^{\pU}$. In contrast, the bias of our bias adjusted estimator, $\hat{\lambda}_{\text{\tiny H,$\mathcal{D}$iag}}^{\pU}$, is as low as that of the benchmark $\hat{\lambda}_{\text{\tiny $\mathcal{I}$,$\mathcal{D}$iag}}^{\pU}$. In the top center panel of Figure \ref{bias_sd_trade_off} we see that the variances of all three estimators are very similar, and decrease as $K$ increases. The decrease is because on average larger $K$ corresponds to more overlap between $p$ and the calibrated $\phi_{\mix}$, and
thus more overlap between $\tilde{p}$ and $\phi$. The top right panel of Figure \ref{bias_sd_trade_off} shows the RMSE of the estimators which is similar for $\hat{\lambda}_{\text{\tiny H,$\mathcal{D}$iag}}^{\pU}$ and $\hat{\lambda}_{\text{\tiny $\mathcal{I}$,$\mathcal{D}$iag}}^{\pU}$, but much larger for $\hat{\lambda}_{\text{\tiny $\mathcal{D}$,$\mathcal{D}$iag}}^{\pU}$ because of its large bias.

\begin{figure}[t]
\begin{center}
\includegraphics[width=0.7\textwidth]{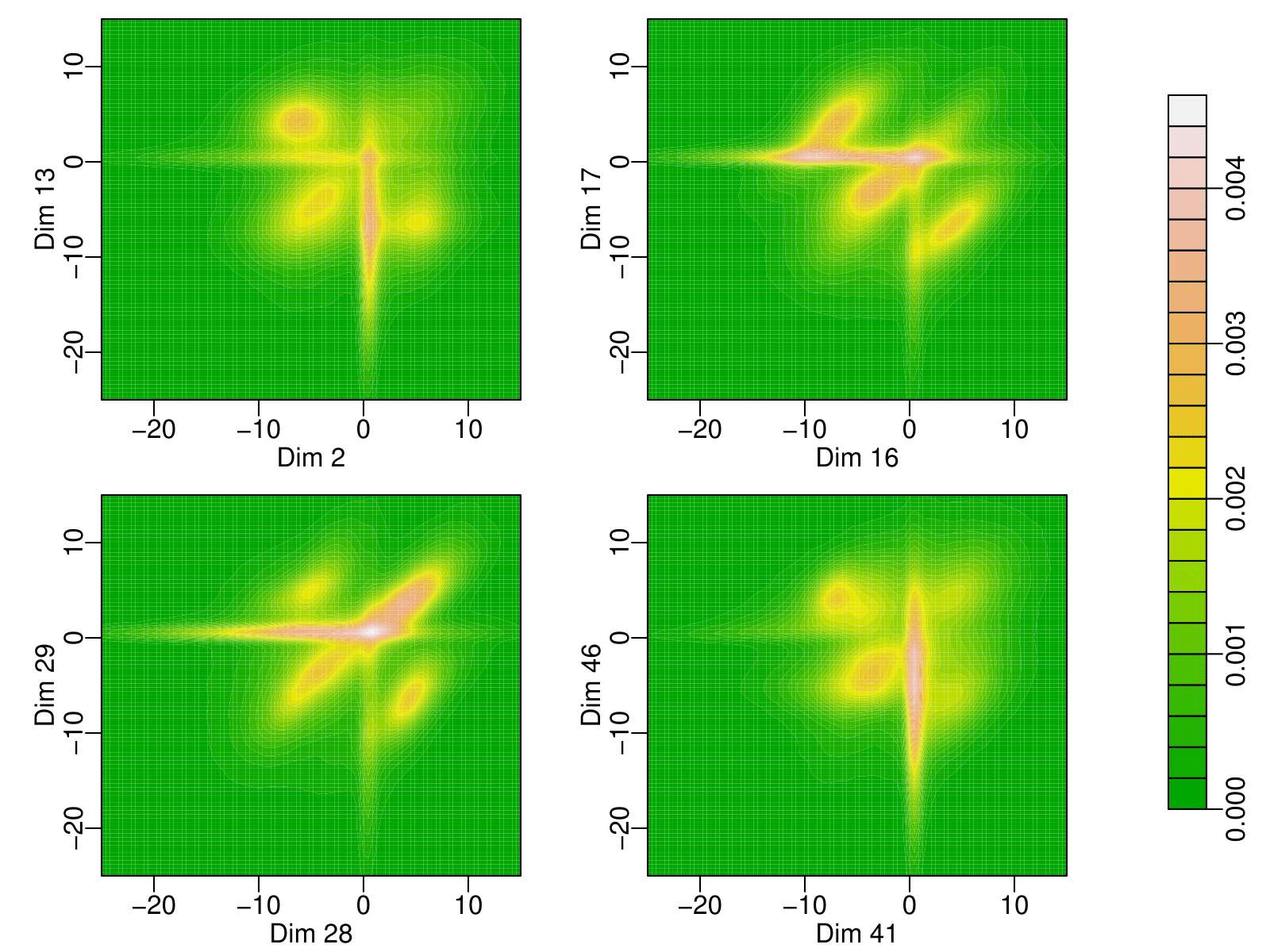}
\caption{\label{fig::data_illustration}\small{Contours of the density $p$ projected onto different pairs of dimensions.}}
\end{center}
\end{figure}

\begin{comment}
\begin{figure}[t]
 \renewcommand{\arraystretch}{0.28}
 \setlength{\tabcolsep}{2pt}
\includegraphics[width=0.8\textwidth,trim=27mm 114mm 24mm 106mm,clip]{fig22.pdf}
\caption{\label{fig::choosing_L}\small{The impact of $L$ on $T_\ttEM$ (left), $T_\opt^\pX$ (middle), and the RMSE of
$\hat{\lambda}_\opt^\pX$ (right) in the 50-dimensional example. Black lines: $K=5$; red: $K=25$; green: $K=50$.
When $K=50$, we can only take $L/K$ up to 100, because $L\leqslant n/2$.}}
\end{figure}
\end{comment}

The bottom row  of Figure \ref{bias_sd_trade_off} shows similar results to those discussed above, but in the case where the covariance matrices of the components of $\phi_{\mix}$ are not constrained to be diagonal. In this setting, we denote the three estimators by  $\hat{\lambda}_{\text{\tiny $\mathcal{D}$,Full}}^{\pU}$, $\hat{\lambda}_{\text{\tiny H,Full}}^{\pU}$, and
$\hat{\lambda}_{\text{\tiny I,Full}}^{\pU}$. The results broadly match those in the top row of Figure \ref{bias_sd_trade_off}, except that the bias (bottom left panel) and RMSE (bottom right panel) of $\hat{\lambda}_{\text{\tiny $\mathcal{D}$,Full}}^{\pU}$  are even larger than those of $\hat{\lambda}_{\text{\tiny $\mathcal{D}$,$\mathcal{D}$iag}}^{\pU}$. This is because with full covariance matrices, we have significantly more parameters to be estimated, and hence more substantial over-fitting bias.  However, Figure~\ref{bias_sd_trade_off} clearly shows that our method removes the adaptive bias regardless of its magnitude, and differences between using full and diagonal covariance matrices  when fitting $\phi_{\mix}$ are minor (compare the dashed lines in the top and bottom panels). Since fitting $\phi_{\mix}$ with diagonal covariance matrices is computationally much less expensive, $\hat{\lambda}_{\text{\tiny H,$\mathcal{D}$iag}}^{\pU}$ achieves better RMSE per CPU second than $\hat{\lambda}_{\text{\tiny H,Full}}^{\pU}$; see Appendix \ref{app:cov} for further demonstration. Hence, from hereon we always use diagonal covariance matrices and the estimation strategy in Section \ref{adaptive_bias} and denote the final estimator by $\hat{\lambda}^{\pX}_{\alpha} = \dfrac{1}{2}
\left(\hat{\lambda}^{\pX}_{\alpha,1}+ \hat{\lambda}^{\pX}_{\alpha,2}\right)$, where $\mathcal{X}=U$ or ``mix". Recall that ``mix" refers to the ordinary bridge sampling estimator using $p$ and $\phi_\mix$, i.e., the logarithm of (\ref{equ::mixture_bridge_sampling}).

\begin{figure}[t]
 \renewcommand{\arraystretch}{0.28}
 \setlength{\tabcolsep}{2pt}
\begin{center}
\includegraphics[width=0.8\textwidth,trim=27mm 115mm 24mm 106mm,clip]{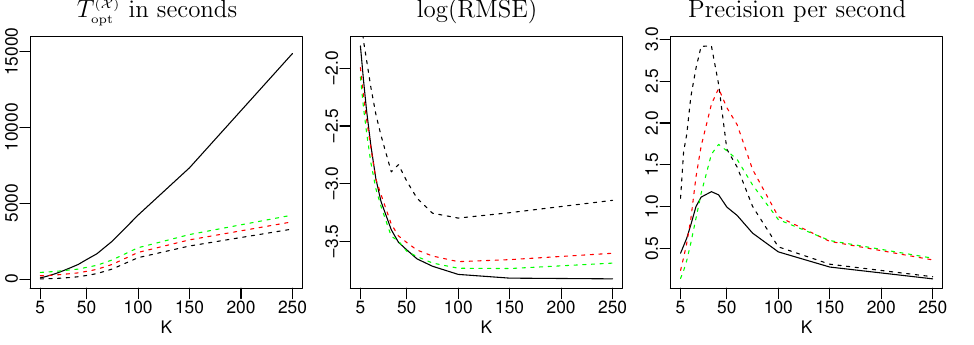}
\caption{\label{fig::50D_PpS_156}\small{The total computational cost (left), the log(RMSE) (middle),
and the $PpS$ (right) of $\hat{\lambda}_{\opt}^\pU$ (solid lines, $m=n$) and
$\hat{\lambda}_{\opt}^\pmix$ (dashed lines) with $m=n$ (black), $16n$ (red), and $32n$ (green).
}}
\end{center}
\end{figure}

Next we consider a 50 dimensional example. For this example, $p$ is a mixture of 30 distributions, including Normal distributions, \textit{t}-distributions (including Cauchy distributions), and multivariate distributions with gamma and/or exponential marginal distributions
and Normal copulas.
The four 2-D projection contour plots of $p$ in
Figure \ref{fig::data_illustration} show
the density has very long tails and is quite skewed in some directions.
Evaluating $p$ is about 700 times more costly than evaluating   $\phi$ (the auxiliary density).
The simulation results are based on $10^4$ replications,
and
in each replication,  $n=10^4$ samples were drawn from $p$.

Figure \ref{fig::50D_PpS_156} shows the total computational cost, the RMSE, and the
$PpS$ (``precision per second", defined as the reciprocal of RMSE$\times$CPU seconds)
of $\hat{\lambda}_{\opt}^\pX$.
As in the 10-dimensional example, the RMSE decreases as $K$ increases up to $n/100$, and when
$K>n/100$, the mixture model overfits the data (i.e., the draws from $p$), resulting in a slight increase in the
RMSE of $\hat{\lambda}_{\opt}^\pmix$. On average,
log(RMSE) of $\hat{\lambda}_\opt^{\pU}$ is about 60\% of that of $\hat{\lambda}_\opt^{\pmix}$, but
the computational cost of $\hat{\lambda}_\opt^{\pU}$ is
4.7 times that of $T^\pmix_\opt$,
so in terms of the $PpS$,
$\hat{\lambda}_\opt^{\pmix}$ is superior to $\hat{\lambda}_\opt^{\pU}$.
In addition,   for large $K$,
when we increase $m$ from $n$ (black lines) to $16n$ (red) and $32n$ (green), the total computational cost of $\hat{\lambda}_\opt^\pmix$
increases only by a small fraction, but the gain in statistical efficiency is substantial. Thus, in this illustration $\hat{\lambda}_\opt^{\pmix}$ is preferred to $\hat{\lambda}_\opt^{\pU}$. However, such preferences vary with machines and implementations, because we have not optimized the function evaluation routines or other aspects of the code. Furthermore, the relatively high computational cost of $\hat{\lambda}_\opt^{\pU}$ seen in Figure \ref{fig::50D_PpS_156} is partly due to obtaining the Warp-U transformed draws $\{\tilde{\omega}_1,\dots,\tilde{\omega}_n\}$, which does not require any target evaluations. %Target evaluations are only needed when evaluating $q$ or its transformed version $\tilde{q}$ given in (\ref{equ::Warp-U-transformed-density}).0
In other scenarios, evaluations of $q$ may dominate the computational cost more, and then computational efficiency would depend mostly on the number of target evaluations.   In particular, in such cases, increasing $m$ from $n$ to say $16n$ would represent an 8.5 (i.e, $(16+1)/2)$) multiplicative increase in computation, as opposed to the relatively modest increase seen in the left panel of Figure \ref{fig::50D_PpS_156}, and therefore the computational cost of $\hat{\lambda}_\opt^{\pmix}$ would be more similar to that of $\hat{\lambda}_\opt^{\pU}$ for a given log(RMSE).  %the most important measure of efficiency would be the number of target evaluations needed to achieve a certain precision.
We consider measuring computational efficiency by the number of target evaluations in Section \ref{sec:comparison}, and find $\hat{\lambda}_\opt^{\pmix}$ and $\hat{\lambda}_\opt^{\pU}$ to be closely comparable. Opportunities for reducing the computational cost associated with Warp-U bridge sampling will be discussed in Section \ref{sec:discussion}.
%On the other hand, obtaining the smallest possible RMSE may be of primary concern and in our studies  this is consistently achieved by $\hat{\lambda}_\opt^{\pU}$ with large $K$, at least in the case where $n$ is fixed.  In practice, the preference for $\hat{\lambda}_\opt^{\pmix}$ or $\hat{\lambda}_\opt^{\pU}$ is likely to depend on the the level of accuracy required and the availability of parallel computing resources.

%Figure \ref{fig::choosing_L} shows the impact of increasing $L/K$ on $T_\ttEM$ (left), $T_\opt^\pX$ (middle), and the log(RMSE) (right) of estimators with $K=5$ (black lines), $25$ (red), and 50 (green). Consistent with Figure \ref{fig::10D_L_over_K}, as $L/K$ increases up to 50, the statistical efficiencies of the estimators improve considerably, but as we continue to increase $L/K$, the slope of the curves of log(RMSE) become very gradual. Hence this example also supports  $L=\min(50K,n/2)$.

%%%%%%%%%%%%%%%%%%%%%%%%%%%%%%%%%%%%%%%%%%%%%%%%%%%%%%
\section{Warp-U extension of Generalized Wang-Landau method}\label{sec:comparison}

%Whereas the theorericalSo far we have considered the scenario in which we have $n$ i.i.d. samples $\omega_1,\dots,\omega_n \overset{}{\sim} p$ at our disposal. In practice, the samples may instead be correlated samples obtained via one of the many Markov chain MC (MCMC) methods available, such as Metropolis-Hastings, Hamiltonian MC, or parallel tempering. However, the correlation between samples can be substantially reduced by thinning, i.e., running the MCMC algorithm for multiple steps between each recorded sample. Thus, the methods described so far also apply when MCMC algorithms are used to obtain samples from $p$. %Thinning to obtain approximately independent samples is required by many MC estimation approaches and so is not a disadvantage of (warp) bridge sampling. %Many MC estimation approaches rely on independent samples from the target so thinning is not a disadvantage of bridge sampling and warp would not be %The impact of correlation can also be accounted for by computing an effective sample size $n_{eff}$ to use in place of $n$ in the bridge sampling estimator, {\color{red} see \dots XL?}

As we emphasized earlier, bridge sampling is applicable to general MCMC settings. We therefore do not need the draws to be independent, as long as they are from the target $p$, or at least in the long run.
There are cases, however, where  the available draws are from a distribution that is known to be different from the target $p$. Indeed, some of the most promising approaches for estimating normalizing constants combine the tasks of sampling and estimation into one coherent algorithm, but often do not directly sample from $p$.  %Many of these combined approaches are forms of the adaptive importance sampling methods mentioned in Section \ref{sec:introduction}. The success of combined sampling and estimation algorithms is intuitive in the sense that an algorithm which uses the current estimation performance to inform future sampling should produce better estimates than approaches which complete all the sampling first and only then consider estimation.
An important case of such a combined approach is the generalized Wang-Landau (GWL) algorithm proposed by \citet{liang2005generalized}. GWL is particularly useful in the current context because of its ability to efficiently sample from highly multi-modal distributions.  We therefore take it as a benchmark, and illustrate how it can be combined with Warp-U bridging sampling to obtain an improved estimator for normalizing constants. More generally,  our strategy of incorporating a Warp-U bridge sampling step can be tried on other algorithms that combine sampling and estimation of normalizing constants. %can be combined with other similar algorithms to offer in a similarly improve the estimatother algorithms%to compare against. The comparison is not direct because our proposed method only regards how to {\it use} samples once they are obtained, and says nothing about how to obtain samples in the first place.\footnote{{\color{red}However, future work is to investigate the potential gains to be made by applying Warp-U in sampling algorithms, see Section \ref{sec:sampling} and Appendix \dots for discussion of this idea.}}  Nonetheless, we will see that there is a substantial benefit to integrating  our Warp-U bridge sampling method into GWL, and it is likely that Warp-U can improve other algorithms in a very similar way.

\subsection{GWL algorithm}\label{sec:gwl}

We begin by briefly describing GWL, which shares some similarities with other energy based methods such as the equi-energy sampler  \citep{kou2006equi}. Suppose that we want to compute the integral $\int_Sq(\omega)\measure(\ud\omega)$, for some unnormalized density $q$ and bounded region $S$. Typically $S$ is the region over which $q$ has non-negligible density, i.e.,  $q(S)\approx 1$.  We divide $S$ into $r$ subregions $S_1,\dots,S_r$, which are defined by target energy bins; that is, within each
$S_i$, the energy level, defined by $-\log q$,  is roughly the same. %Sampling is then performed using a modified Metropolis--Hastings algorithm.
Let the current estimate of the integral $\int_{S_i} q(\omega)\measure(\ud\omega)$ be denoted by $\hat{g}(S_i)$, and set the initial estimate to be $\hat{g}(S_i)=1$, for $i=1,\dots,r$. %Each time the subregion $E_i$ is sampled a small constant $\delta$ is added to the estimate  $\hat{g}(S_i)$.
GWL takes the inputs $n_0$, $\delta_0$, and $T$ (e.g., $n_0=1000$, $\delta_0=e-1\approx 1.718$, $T=25$) and proceeds as detailed below. Note that, \citet{liang2005generalized} introduces an additional tuning parameter, which is not needed when the goal is to estimate normalizing constants, and therefore it is not included here.
\vspace{0.2cm}

\noindent {\bf GWL algorithm.}\\
\noindent For stage $t=1,\dots,T$:
\vspace{-\topsep}
\begin{enumerate}
\setlength{\itemsep}{-0.3cm}
    \item Set $n_t=n_{t-1}(1.1)^{t-1}$, $\delta_t=\sqrt{1+\delta_{t-1}}-1$, and $\hat{g}(S_i)^{(t,1)}=\hat{g}(S_i)^{(t-1,n_{t-1})}$, for $i=1,\dots,r$.
    \item For $k=1,\dots, n_t$ do the following:
    \vspace{-\topsep}
    \begin{enumerate}[(i)]
    \setlength{\itemsep}{-0.3cm}
     \setlength{\topskip}{-2cm}
    \item Use a Metropolis-Hastings step, with proposal density $h$, to
draw a sample $\omega$ from
    the current target density
    \vspace{-0.3cm}
    \begin{align}
        \psi^{(t,k)}(\omega) \propto \sum_{i=1}^r \frac{q(\omega)}{\hat{g}^{(t,k)}(S_i)}I(\omega\in S_i).\label{eqn:intermediate}
    \end{align}
    The $(t,k)$ superscripts indicate that the current target and the estimate of $g(S_i)$, for $i=1,\dots,r$, are updated in each iteration within each stage.\\
\item Update $\hat{g}^{(t,k)}(S_{I_\omega})$ to $(1+\delta_t)\hat{g}^{(t,k)}(S_{I_\omega})$, where $I_\omega$ is the index such that $\omega \in S_{I_\omega}$, i.e., a regional mass estimate $\hat{g}^{(t,k)}(S_i)$ is increased only if $S_i$ contains the draw $\omega$.
\end{enumerate}
\end{enumerate}%Since sampling can in theory continue indefinitely, the estimates $\hat{g}(S_i)$ continue to increase and are only determined up to a common constant $c$ (which changes with the number of Metropolis--Hastings iterations). However, there is a way to estimate $c$, which we discuss in the next paragraph. Moving on,
%For simplicity suppose the proposal density is symmetric. Then, %letting $I_x$ be such that $x\in S_{I_x}$,
%if the current sample $\omega$ is in subregion $S_{I_\omega}$, the acceptance probability of a new sample $\omega^*$ in subregion $S_{I_{\omega^*}}$ is given by $\min\left\{\frac{\hat{g}(S_{I_\omega})p(\omega^*)}{\hat{g}(S_{I_{\omega^*}})p(\omega)},1\right\}$.
%The approach in \citet{liang2005generalized} is to run this algorithm in multiple stages, where each new stage corresponds to a decrease in $\delta$ and an increase in the number of iterations. This method leads

It should be clear from the description above that, in the limit, GWL samples the subregions $S_1, \ldots, S_r$ with equal probability, and within each $S_i$, it samples according to $q$. Therefore its stationary distribution is not the targeted $q$, but what can intuitively be described as a ``re-distributed" $q$ that equalizes the masses of the energy bins:
\begin{equation}
\psi(\omega)=\sum_{i=1}^r \frac{q(\omega)}{g(S_i)}I(\omega\in S_i).
\end{equation}
\citet{liang2005generalized} verified this convergence assuming that the intermediate densities of (\ref{eqn:intermediate}) can be sampled from exactly. Practically, we can sample from them approximately, say by repeating the Metropolis-Hastings step many times between each update of the estimate of $g(S_i)$, for $i=1,\dots,r$. However, \citet{liang2005generalized} used only one Metropolis-Hastings update in his illustrations, and we follow this practice (but with an ideal proposal, see Appendix \ref{app:gwl_details}). His proof also assumed $n_t$ grows sufficiently fast with $t$, but the multiplicative factor $1.1$ suggested in  \citet{liang2005generalized} may not always be adequately large, though it seems computationally problematic to increase it much further.

Since GWL samples the subregions uniformly, the final estimators $\hat{g}(S_i)=\hat{g}^{(T,n_T)}(S_i)$  estimate the integrals $g(S_i)$ only up to a common constant, denoted $A$, which depends on $\delta_0$ and the number of iterations made at each stage of the algorithm. To estimate the log normalizing constant of $q$, namely $\lambda=\log(c)$, we must remove $A$, which can be done by running GWL with a modified version of $q$ as we now explain. Choose $S_2,\dots,S_r$ to be such that $\left(\bigcup_{i=2}^rS_i\right)^c$ has negligible mass under $q$, and choose $S_1 \subset \left(\bigcup_{i=2}^rS_i\right)^c$ such that its volume $|S_1|$ is finite and known \citep{liang2005generalized}. Next, run GWL with $q$ replaced by
\begin{align*}
    q_{\text{mod}}(\omega) = \left\{\begin{array}{cc}q(\omega) & \mbox{for } \omega \in \bigcup_{i=2}^rS_i\\
    \frac{1}{|S_1|} & \mbox{for }\omega\in S_1\\
    0 & \mbox{otherwise.}\end{array}\right.
\end{align*}
The basic idea is that on $S_1$ we can treat $q$ as uniform since its actual distribution contributes little to the normalizing constant of $q$.
Lastly, for each $S\in\{S_2,\dots,S_r\}$, \citet{liang2005generalized} %consistently
estimated the integral $\int_S q(\omega)d\omega$ by  $\hat{g}(S)/\hat{g}(S_{1})$. Assuming convergence of GWL, a consistent estimate of $\lambda$ is thus given by $\hat{\lambda}_{GWL}=\log(\sum_{i=2}^r\hat{g}(S_i)/\hat{g}(S_{1}))$.

The key strengths of GWL are its adaptive nature and that it exhibits good mixing properties even for multi-modal targets, the latter property being a benefit of asymptotic uniform sampling across energy bins. In practice, a limitation of the algorithm  is that  the MSE of $\hat{\lambda}_{GWL}$ is bounded below for fixed $n_0$ and the specified geometric growth in $n_t$, as can be seen in the top left panel of Figure \ref{gwl_results} (discussed in Section \ref{sec:gwl_results}). (\citet{liang2007stochastic} attempted to mitigate this phenomenon, but the convergence properties of the updated algorithm again may not be ideal, and further developments are still being made; see for example \citet{jacob2014wang}.) Here we simply view GWL as a related method to compare against and combine with. For these purposes the lower bound on the convergence of $\hat{\lambda}_{GWL}$  does not play a large role because the number of target evaluations we allow is approximately equal to or lower than the number required by GWL to achieve its minimum MSE.
For further details of GWL the reader is referred to \citet{liang2005generalized}, \citet{liang2007stochastic},  \citet{bornn2013adaptive}, and \citet{jacob2014wang}.

\subsection{Combining GWL with Warp-U}\label{sec:gwl+warpu}

%We now consider the situation where GWL has been run for 10 stages, we think it is near convergence, and we can decide  whether to use the computation needed for an eleventh stage to continue running the algorithm (GWL-only) or to instead make use of Warp-U bridge sampling  to obtain $\hat{\lambda}^{\pU}_{\opt}$ (GWL+Warp-U). Draws from $p$ are required to implement the latter, but samples can be obtained from the first 10 stages of the GWL run without any additional target evaluations. In particular, to obtain a sample approximately from $p$ we first sample $k$ from $1,\dots,r$ according to the weights $w_i=\hat{g}(S_i)/\sum_{j=1}^r\hat{g}(S_j)$, for $i=1,\dots,r$, and then uniformly select one of the samples in subregion $S_k$ recorded during the first 10 stages of GWL. Since these samples require no new target evaluations, the only  new target evaluations required for Warp-U bridge sampling is the $mK$  needed to evaluate $\tilde{p}$ at the $m$ samples from $\phi$.

Consider a situation where GWL has been run for $T^* < T$ stages. We suspect that it is near convergence, and want to determine if we can reduce the MSE by using the computation to only complete the remaining $T-T^*$ stages (the GWL-only approach) or to make the use of Warp-U bridge sampling to obtain the estimator $\hat{\lambda}^{\pU}_{\opt}$ (the GWL+Warp approach). Draws from our target $p$ are required to implement the latter approach, but can be obtained from the GWL run without any additional target evaluations. To see this, first let $G_i$ denote the set of samples collected from subregion $S_i$ during the $T^*$ stages of the GWL run, for $i=1,\dots,r$.  With this notation, we propose the following addition to GWL.
\vspace{0.2cm}

\noindent{\bf Warp-U Addition.}
\noindent
\vspace{-\topsep}
\begin{enumerate}
\setlength{\itemsep}{-0.3cm}
\item For $l=1,\dots,n$, repeat the following two steps:
\vspace{-\topsep}
    \begin{enumerate}[(i)]
\setlength{\itemsep}{-0.3cm}
    \item Sample a subregion index $k\in\{2,\dots,r\}$ using the probabilities $b_i \propto \hat{g}(S_i)1_{\{G_i\not=\emptyset\}}$, for $i=2,\dots,r$.\\\vspace{-0.3cm}
    \item With uniform sampling, select a sample $\omega_l \in G_k$, i.e., select  one of the samples in subregion $S_k$ collected by GWL. \\\vspace{-0.3cm}
\end{enumerate}
    \item Apply Warp-U bridge sampling with $\{\omega_1,\dots,\omega_n\}$ and $m$ draws from $\phi$ to obtain the estimate $\hat{\lambda}^{\pU}_{\opt}$.
\end{enumerate}
The first step obtains draws from $p$ restricted to $\bigcup_{i=2}^rS_i$, and the second step applies Warp-U bridge sampling using these draws. The indicator $1_{\{G_i\not=\emptyset\}}$ in $b_i$ indicates that
we sample only  regions from which there are samples during the GWL run. %When this Warp-U addition is run after GWL we refer to the whole algorithm as the {\it GWL+Warp-U} approach, and if it is not run then we refer to the algorithm as the {\it GWL-only} approach.
Since $q(\omega_l)$, for $l=1,\dots,n$, has already been evaluated during the GWL run, the only  new target evaluations required for the above Warp-U addition are the $mK$  needed to evaluate $\tilde{q}$ at the $m$ draws from $\phi$. In step 2 above, we could alternatively apply standard bridge sampling to $p$ and $\phi_{\mix}$ to obtain $\hat{\lambda}^{\mix}_{\opt}$, as described in Section \ref{subsection::simulation_known_mixture}. We refer to this alternative approach as GWL+BS.

\subsection{Illustration of the GWL+Warp-U algorithm}
\label{sec:gwl_results}

\begin{figure}[t]
\begin{center}
\includegraphics[width=0.9\textwidth,trim=0mm 5mm 0mm 0mm,clip]{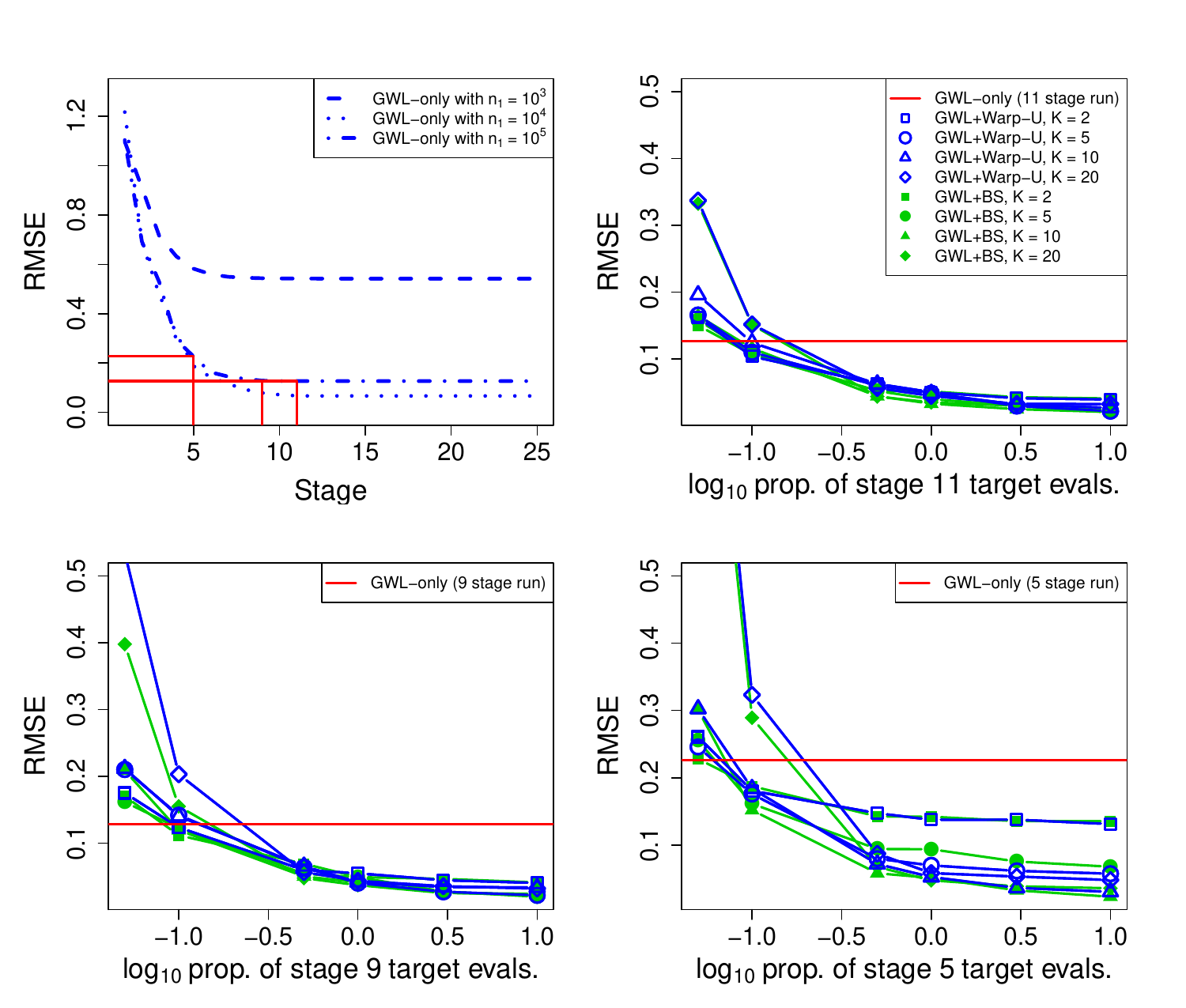}
%\caption{Graphical illustration of Warp-II and Warp-III transformation.}
\caption{\label{gwl_results}\small {The top left panel shows the RMSE of $\hat{\lambda}_{GWL}$ at each of 25 stages, for $n_0=10^3,10^4,10^5$. The solid lines indicate the results for 11-stage, 9-stage, and 5-stage runs of the GWL-only algorithm with $n_0=10^4$. The top right panel shows the RMSE of the GWL+Warp-U estimator (hollow symbols) and the GWL+BS estimator (solid symbols), where the initial GWL run is for 10 stages. The number of target evaluations used by the Warp-U or ordinary bridge sampling step is given on the {\it x}-axis as a  proportion of the target evaluations that would be needed for an eleventh stage of GWL (on a $\log_{10}$ scale). The different shapes correspond to different settings of $K$. %The solid horizontal line shows the $\log_{10}$ MSE when the GWL is run for 11 stages. All results are based on 100 repetitions and the corresponding uncertainties are relatively small.
The bottom left and right panels show similar results where the initial GWL run is for 8 and 4 stages, respectively. }}
\end{center}\end{figure}

We consider the 25 skewed-{\it t} mixture example from Section \ref{adaptive_bias}. We run the GWL-only algorithm for $T=25$ stages, and again set $\delta_0=e-1$. We try $n_0=10^3,10^4,10^5$ and find that larger $n_0$ requires more stages for the estimator $\hat{\lambda}_{GWL}$ to converge but leads to lower RMSE; see the top left panel of Figure \ref{gwl_results}. %The total number of target evaluations across the 25 stages  for these values of $n_0$ are approximately $10^5$, $10^6$, and $10^7$, respectively.
We choose $n_0=10^4$ as a compromise between RMSE and computational cost. The value of $n_0$ is not the target of our comparisons, nor is the choice of the proposal density $h$ or the partition $S_1,\dots, S_r$. We therefore
use our knowledge of the true target $p$ to configure these components to favor the GWL-only algorithm; see Appendix \ref{app:gwl_details} for details.

%The left panel of Figure \ref{gwl_results} shows $\log_{10}$ of the mean squared error (MSE) for GWL estimator of the normalizing constant of $p$ (the true value is 1) at each of the 25 stages. The {\it x}-axis shows $\log_{10}$ of the cumulative number of target evaluations used. The dotted and dashed lines indicate the results at stage 4 and stage 5. These results are based on 100 runs of the algorithm.

The solid lines in the top left panel of Figure \ref{gwl_results} indicate the RMSE when the GWL-only algorithm (with $n=10^4$) is run for 5, 9, and 11 stages, and we see the RMSE is stabilized after stage 9. For reference, the same three RMSE values are indicated by a horizontal line in the top right, bottom left, and bottom right panels of  Figure \ref{gwl_results}, respectively. Running GWL for more than 9 stages does not improve the RMSE, but we can improve it using the Warp-U Addition detailed in Section \ref{sec:gwl+warpu}, even if we do not increase the overall computational cost. In the top right panel of Figure \ref{gwl_results}, the hollow symbols show the RMSE for the GWL+Warp-U estimator $\hat{\lambda}^{\pU}_{\opt}$, in the case where the GWL component was run for only 10 stages, thereby saving the computation needed for an $11^{\text{th}}$ stage to be used for Warp-U sampling.  Of course, we may decide we do not need to use all of the saved computation. The $x$-axis of Figure \ref{gwl_results} gives the number of target evaluations we use for the Warp-U step, in units of $\log_{10}$ of the proportion of target evaluations needed for an $11^{\text{th}}$ stage of GWL, i.e., at $0$ on the $x$-axis the computational cost of the Warp-U sampling matches that of an $11^{\text{th}}$ stage. The different hollow shapes correspond to different values of $K$ (number of mixture components). For comparison, the horizontal line indicates the RMSE under the GWL-only algorithm after 11 stages (for which the $x$-axis is irrelevant). At $0$ on the $x$-axis the GWL-only and GWL+Warp-U methods use the same number of target evaluations, but the GWL+Warp-U method yields substantially lower RMSE for all four values of $K$ (2, 5, 10, and 20). Indeed, the RMSE obtained is even lower than that achieved by the GWL-only algorithm with $n_0=10^5$ after 11 (or 25) stages, which uses a factor of 10 more target evaluations than used by the  GWL+Warp-U runs (with $n_0=10^4$). %When we apply Warp-U with only half the number of target evaluations needed for an eleventh stage of GWL (around $-0.3$ on the {\it x}-axis) we still see a large reduction in RMSE compared with running an eleventh stage.
This example suggests that applying Warp-U when GWL is at or close to convergence offers a way to substantially lower RMSE with only the number of target evaluations required to run one more stage of GWL (or even fewer). The improvements offered by GWL+Warp-U are thus almost free because we can stop GWL one stage early, and in any case the number of target evaluations required by the Warp-U part is relatively low. %Even if the Warp-U part is viewed as extra computation, the number of target evaluations it requires is relatively low compared with the total number used by GWL.

%Even for the case where only a tenth of the target evaluations needed for a fifth stage of GWL are used ($-1$ on the {\it x}-axis), GWL+Warp-U still offers large improvements over the GWL only mention for $K=5,10,20$.

For further comparison, the solid shapes in the top right panel of Figure \ref{gwl_results} show results for the GWL+BS method, i.e, where the two bridge sampling densities are $p$ and $\phi_{\mix}$. Note that, at any given point on the {\it x}-axis the number of target evaluations is the same for the GWL+BS and GWL+Warp-U algorithms, which is achieved by setting $m=n(2K-1)$ in the GWL+BS algorithm and $m=n$ in the GWL+Warp-U algorithm. We see that GWL+BS again achieves substantially lower RMSE than is obtained by simply running an $11^{\text{th}}$ stage of the GWL-only algorithm. The RMSE under the GWL+BS method is also seen to be marginally lower than that under the GWL+Warp-U method, but now that we have controlled the number of target evaluations, we see the performance is very similar. % For a shorter initial run of GWL (bottom right panel of Figure \ref{gwl_results}) we see that     However, the MSE under the GWL+BS method is always higher than that under the GWL+Warp-U approach, again illustrating the additional gains made possible by the Warp-U transformation in the context of multi-modal $p$.

Naturally, Warp-U or standard bridge sampling can be applied after any number of GWL stages; we do not have to wait until we are sure of convergence. To investigate this, we ran the GWL+Warp-U and GWL+BS algorithms again but where the initial GWL run was shorter. The bottom left and bottom right panel of Figure \ref{gwl_results} show results in the case where the initial GWL run was 8 and 4 stages, respectively. These results are qualitatively similar to before, except that GWL+Warp-U occasionally performs slightly better than GWL+BS (e.g., when $K=5$ in the bottom right panel).  The main constraint on when Warp-U or standard bridge sampling can be used is that GWL needs to have been run for long enough to have explored regions where $p$ has high density.  Otherwise, weighted re-sampling of the GWL samples will not yield approximate draws from $p$. This is not a major issue in practice because, % neither GWL or any method that uses its output is likely to produce a good normalizing constant estimate if $p$ has not been properly explored.
as can be seen in the top left panel of Figure \ref{gwl_results}, GWL substantially reduces RMSE in its initial stages, and it is likely to be the latter stages where applying Warp-U or standard bridge sampling is  most appealing.

\section{Strategies for Making Further Improvements}
\label{sec:discussion}

We have seen that stochastic Warp-U transformations can transform a multi-modal density $q$ into an approximate Normal density (or some other convenient auxiliary density $\phi$). The transformed density $\tilde{q}$ has the same normalizing constant $c$ as the original target density $q$. Furthermore, since the overlap between $\tilde{q}$ and the base density $\phi$ is typically large, performing bridge sampling with the transformed draws $\{\tilde{\omega}_1,\dots,\tilde{\omega}_n\}$ produces an accurate estimate of $c$. In particular, Theorem 1 implies that the accuracy is equal to or greater than that obtained if we perform bridge sampling using the original draws $\{\omega_1,\dots,\omega_n\}$ and the densities $q$ and $\phi_\mix$. Thus, in terms of statistical efficiency, Warp-U transformations are always beneficial, especially when the target density is multi-modal. However, there are still a number of important directions for further developments, especially regarding computational efficiency and identifying a good approximating mixture distribution $\phi_{\mix}$, which we now discuss. % We now discuss these opportunities for further work.

\subsection{Reducing the Cost of Transformed Density Evaluations}

From (\ref{equ::Warp-U-transformed-density}), it is seen that the density $\tilde{q}$ is at least $K$ times more expensive to evaluate than $q$. Consequently, in Figure \ref{gwl_results} we saw that despite its greater statistical efficiency, Warp-U bridge sampling performs similarly to ordinary bridge sampling with  $q$ and $\phi_\mix$, after accounting for the number of target evaluations needed, i.e., evaluations of $q$.  %In practice, the case of expensive $q$ evaluations is of most interest, but in other situations where $q$ is not expensive enough to dominate the computational cost, %In Figure in Section we saw that when $q$ is not expensive enough to dominate computational cost, then the   in some cases the precision per second The most important case in practice is when $q$ is
When evaluating $q$ is inexpensive, ordinary bridge sampling can perform substantially better than Warp-U bridge sampling, in terms of precision per CPU second, e.g., see Figure \ref{fig::50D_PpS_156}. The greater efficiency of ordinary bridge sampling in such cases is due to the many multivariate Normal evaluations required by Warp-U bridge sampling. These evaluations are not prohibitively expensive and in practice will often be of secondary concern to $q$ evaluations. Cases where $q$ is inexpensive to evaluate  are generally of less interest because many methods will be practical in this scenario. Thus, we are primarily concerned with the statistical efficiency of Warp-U as a function of $q$ evaluations.

We therefore need to find ways to reduce the computational cost of evaluating $\tilde{q}$.
Direct density approximation is unlikely to be fruitful in high dimensions. A plausible strategy is to randomly select only some of the $K$ terms in (\ref{equ::Warp-U-transformed-density}) to evaluate, as is done in the mixture sampling method of \citet{elvira2019generalized}; they randomly partitioned the set of sampling densities and then used these partitions in computing importance weights. Specifically, the importance weight for a sample $\omega$ drawn from density $p_s$ is set to be  the reciprocal of the average of the sampling densities in the same partition as $p_s$. % the true density that $\omega$ was drawn from. as the true  single partition when computing each importance weight. %  for reducing the computational cost of Warp-U sampling is to approximate the density $\tilde{q}(\tilde{w};\bz)$ in (\ref{equ::Warp-U-transformed-density}).
We could apply a similar approach by viewing the terms $c\tilde{p}^{(k)}(\tilde{\omega})=cp^{(k)}(S_k\tilde{\omega}+\mu_k)$) in (\ref{equ::Warp-U-transformed-density})  as the weighted and unnormalized sampling densities.
However, we do not know from which weighted component $c\tilde{p}^{(k)}$ each sample $\tilde{\omega}\sim \tilde{p}$ originated. Furthermore, the $c\tilde{p}^{(k)}$ incorporate unknown weights, so the unweighted mixture components of $\tilde{q}$ are not available and therefore we cannot weight them by their empirical weights, as is required by (generalized) bridge sampling, e.g., the weights $s_i$, for $i=1,2$, in (\ref{optestimator}).  Regarding the unknown sampling component $c\tilde{p}^{(k)}$, it may be sufficient to stochastically impute  the index of the ``true"  component  by drawing  from the conditional distribution $\varpi(\Psi|\omega)$ in (\ref{equ::varpi_theta_omega}). A possible solution to the second difficulty is to use the theoretical component weights incorporated in the $c\tilde{p}^{(k)}$ in the final bridge sampling estimator as opposed to the observed weights. In summary, there are feasible methods for improving the computational efficiency of Warp-U bridge sampling,  but the several potential strategies need to be further explored. %, including the use of a better adapted reference density, discussed below.

%In addition to increasing computational efficiency, it is also of interest to simply reduce computation time which is likely possible through parallel computation. Bridge sampling with the  Warp-U transformed data is embarrassingly parallelizable, because each sample can be treated individually until the final computation of (\ref{optestimator}).
%If a large number of nodes are available then the bottleneck may in fact be the estimation of $\bz$ via the EM algorithm. Thus, a parallel version of EM or a clustering method, such as the $k$-means algorithm, could be especially effective in reducing the computation time of Warp-U bridge sampling estimators.

\subsection{Base and Mixture Distribution Selection}
\label{sec:base_and_mle}

So far we have focused on using the standard Normal as the base distribution $\phi$. But
Theorem 1 implies that
the $f$-divergence between $(\phi_\mix, p)$ is larger than that between $(\phi,\widetilde{p})$ for {\it any} continuous densities $p$ and $\phi$.
For heavy-tailed targets, using a {\it t}-distribution may be more effective for capturing
the mass of $p$, that is, a {\it t}-density may allow us to use a mixture $\phi_{\mix}$ with fewer components and thereby reduce computational costs.
In other contexts, the support of $p$ may be bounded and then a base density with bounded support would be more appropriate. Perhaps in some scenarios it would make sense for $\phi_{\mix}$ to be a mixture of several different base densities, in which case some modifications to the development here would be needed. With a different choice of the base function, a different method for fitting $\phi_{\mix}$ would be required, e.g.,  for the {\it t}-distribution, we could use the approach of  \citet{peel2000robust} to fit $\phi_{\mix}$.
%Nonetheless, we suggest using the diagonal covariance matrix for the components in $\phi_\mix$ and
%using our strategy in Figure \ref{fig::solution_small_bias_variance} to remove the adaptive bias.
%

Secondly, although our approach for fitting the Warp-U parameters  $\bz$ is promising in practice, it is almost certainly not optimal. \citet{kong2003theory} showed that a standard bridge sampling estimator is in fact a maximum likelihood estimator (MLE), and it may be possible to use the same likelihood framework to find an optimal estimator for $\bz$.
More specifically, let $\phi_i$
be the pdf of $\Normal(\mu_i, \Sigma_i)$, for $i=1,\ldots, K$, and
$\phi_\mix=\sum_{i=1}^K\pi_i\phi_i$. %,
%$n_i$ be the number of draws from $\phi_i$, and $m=\sum_{i=1}^Kn_i$.
Then the maximum likelihood estimator of $c$ (with $p=q/c$ as before) identified by  \citet{kong2003theory} is \begin{equation}
\hat{c} = \sum_{i=1}^2\sum_{j=1}^{n_i}\frac{q(w_{i,j})}{n_1\hat{c}^{-1}q(w_{i,j})+n_2\phi_{\mix}(\omega_{i,j})},\label{eqn:mle}
\end{equation}
where $\{\omega_{1,1},\dots,\omega_{1,n_1}\}$ are draws from $p$, and $\{\omega_{2,1},\dots,\omega_{2,n_2}\}$ are draws from $\phi_{\mix}$.
%When  $n_i/m\rightarrow\pi_i$,
The estimator (\ref{eqn:mle}) is the same as  $\hat{c}_{\opt}^{\pmix}=\exp(\hat{\lambda}_{\opt}^{\pmix})$. %The estimator (\ref{eqn:mle}) is derived from a likelihood that treats the baseline measure $\measure(\ud\omega)$ as the unknown, which is natural considering that all integral estimators approximate the baseline measure in some way. \citet{kong2003theory} and \citet{kong2007further} build on this by showing that statistical efficiency  can be further improved, at the cost of computation, by enforcing (true) invariances of $\measure(\ud\omega)$ under selective transformations of $\omega$.
If Warp-U transformations can be correctly incorporated into this likelihood framework then we can use the MLE of $(c,\bz)$ to improve upon our current approach. %\citet{jones_qualifying} provides further insights on how this likelihood formulation might be achieved and identifies some open challenges.
Whether using a maximum likelihood approach or not, it is especially important to investigate further how the number of components $K$ should be chosen.

\subsection{It's Time to Build a Bridge}
The vast majority of the MC literature is about improving MC \textit{sampling} efficiency, that is, how to design MC sampling algorithms most effectively. In contrast,  bridge sampling, with or without warp transformations, is about improving MC \textit{inference} efficiency, that is, how to gain more precision with a given set of MC draws. Adding warp bridge sampling to GWL provides an example of bridging the sampling and analysis approaches, a strategy we believe has much more to offer than the current literature recognizes.  We therefore invite interested readers to join us in laying further foundations for this much needed bridge.

\bibliographystyle{chicago}
\bibliography{warpU_bridge_sampling.bbl} %{citations} %

\begin{thebibliography}{}

\bibitem[\protect\citeauthoryear{Ali and Silvey}{Ali and
  Silvey}{1966}]{ali1966general}
Ali, S.~M. and S.~D. Silvey (1966).
\newblock A general class of coefficients of divergence of one distribution
  from another.
\newblock {\em Journal of the Royal Statistical Society. Series B
  (Methodological)\/}~{\em 28}, 131--142.

\bibitem[\protect\citeauthoryear{Alspach and Sorenson}{Alspach and
  Sorenson}{1972}]{alspach1972nonlinear}
Alspach, D.~L. and H.~W. Sorenson (1972).
\newblock Nonlinear {B}ayesian estimation using {G}aussian sum approximations.
\newblock {\em IEEE Transactions on Automatic Control\/}~{\em 17\/}(4),
  439--448.

\bibitem[\protect\citeauthoryear{Azzalini}{Azzalini}{2011}]{azzalini2011r}
Azzalini, A. (2011).
\newblock R package sn: The skew-normal and skew-{\it t} distributions (version
  0.4-17).
\newblock {\em URL http://azzalini.stat.unipd.it/SN\/}.

\bibitem[\protect\citeauthoryear{Azzalini}{Azzalini}{2013}]{azzalini2013skew}
Azzalini, A. (2013).
\newblock {\em The skew-normal and related families}.
\newblock Cambridge University Press, New York, NY.

\bibitem[\protect\citeauthoryear{Bennett}{Bennett}{1976}]{bennett1976efficient}
Bennett, C.~H. (1976).
\newblock Efficient estimation of free energy differences from {M}onte {C}arlo
  data.
\newblock {\em Journal of Computational Physics\/}~{\em 22\/}(2), 245--268.

\bibitem[\protect\citeauthoryear{Berg and Neuhaus}{Berg and
  Neuhaus}{1991}]{berg1991multicanonical}
Berg, B.~A. and T.~Neuhaus (1991).
\newblock Multicanonical algorithms for first order phase transitions.
\newblock {\em Physics Letters B\/}~{\em 267\/}(2), 249--253.

\bibitem[\protect\citeauthoryear{Bornkamp}{Bornkamp}{2011}]{bornkamp2011approximating}
Bornkamp, B. (2011).
\newblock Approximating probability densities by iterated {L}aplace
  approximations.
\newblock {\em Journal of Computational and Graphical Statistics\/}~{\em
  20\/}(3), 656--669.

\bibitem[\protect\citeauthoryear{Bornn, Jacob, Del~Moral, and Doucet}{Bornn
  et~al.}{2013}]{bornn2013adaptive}
Bornn, L., P.~E. Jacob, P.~Del~Moral, and A.~Doucet (2013).
\newblock An adaptive interacting {W}ang--{L}andau algorithm for automatic
  density exploration.
\newblock {\em Journal of Computational and Graphical Statistics\/}~{\em
  22\/}(3), 749--773.

\bibitem[\protect\citeauthoryear{Ceperley}{Ceperley}{1995}]{ceperley1995path}
Ceperley, D.~M. (1995).
\newblock Path integrals in the theory of condensed helium.
\newblock {\em Reviews of Modern Physics\/}~{\em 67\/}(2), 279--355.

\bibitem[\protect\citeauthoryear{Chen and Tan}{Chen and
  Tan}{2009}]{chen2009inference}
Chen, J. and X.~Tan (2009).
\newblock Inference for multivariate normal mixtures.
\newblock {\em Journal of Multivariate Analysis\/}~{\em 100\/}(7), 1367--1383.

\bibitem[\protect\citeauthoryear{Chen, Tan, and Zhang}{Chen
  et~al.}{2008}]{chen2008inference}
Chen, J., X.~Tan, and R.~Zhang (2008).
\newblock Inference for normal mixtures in mean and variance.
\newblock {\em Statistica Sinica\/}~{\em 18\/}(2), 443--465.

\bibitem[\protect\citeauthoryear{Chib}{Chib}{1995}]{chib1995marginal}
Chib, S. (1995).
\newblock Marginal likelihood from the {G}ibbs output.
\newblock {\em Journal of the American Statistical Association\/}~{\em
  90\/}(432), 1313--1321.

\bibitem[\protect\citeauthoryear{Chib and Jeliazkov}{Chib and
  Jeliazkov}{2001}]{chib2001marginal}
Chib, S. and I.~Jeliazkov (2001).
\newblock Marginal likelihood from the {M}etropolis--{H}astings output.
\newblock {\em Journal of the American Statistical Association\/}~{\em
  96\/}(453), 270--281.

\bibitem[\protect\citeauthoryear{Day}{Day}{1969}]{day1969estimating}
Day, N.~E. (1969).
\newblock Estimating the components of a mixture of normal distributions.
\newblock {\em Biometrika\/}~{\em 56\/}(3), 463--474.

\bibitem[\protect\citeauthoryear{DiCiccio, Kass, Raftery, and
  Wasserman}{DiCiccio et~al.}{1997}]{diciccio1997computing}
DiCiccio, T.~J., R.~E. Kass, A.~Raftery, and L.~Wasserman (1997).
\newblock Computing {B}ayes factors by combining simulation and asymptotic
  approximations.
\newblock {\em Journal of the American Statistical Association\/}~{\em
  92\/}(439), 903--915.

\bibitem[\protect\citeauthoryear{Elvira, Martino, Luengo, and Bugallo}{Elvira
  et~al.}{2015}]{elvira2015efficient}
Elvira, V., L.~Martino, D.~Luengo, and M.~F. Bugallo (2015).
\newblock Efficient multiple importance sampling estimators.
\newblock {\em IEEE Signal Processing Letters\/}~{\em 22\/}(10), 1757--1761.

\bibitem[\protect\citeauthoryear{Elvira, Martino, Luengo, Bugallo,
  et~al.}{Elvira et~al.}{2019}]{elvira2019generalized}
Elvira, V., L.~Martino, D.~Luengo, M.~F. Bugallo, et~al. (2019).
\newblock Generalized multiple importance sampling.
\newblock {\em Statistical Science\/}~{\em 34\/}(1), 129--155.

\bibitem[\protect\citeauthoryear{Gelman, Carlin, Stern, Dunson, Vehtari, and
  Rubin}{Gelman et~al.}{2013}]{gelman2013bayesian}
Gelman, A., J.~B. Carlin, H.~S. Stern, D.~B. Dunson, A.~Vehtari, and D.~B.
  Rubin (2013).
\newblock {\em Bayesian data analysis}.
\newblock CRC Press, Boca Raton, FL.

\bibitem[\protect\citeauthoryear{Gelman and Meng}{Gelman and
  Meng}{1998}]{gelman1998simulating}
Gelman, A. and X.-L. Meng (1998).
\newblock Simulating normalizing constants: From importance sampling to bridge
  sampling to path sampling.
\newblock {\em Statistical Science\/}~{\em 13}, 163--185.

\bibitem[\protect\citeauthoryear{Gronau, Sarafoglou, Matzke, Ly, Boehm,
  Marsman, Leslie, Forster, Wagenmakers, and Steingroever}{Gronau
  et~al.}{2017}]{gronau2017tutorial}
Gronau, Q.~F., A.~Sarafoglou, D.~Matzke, A.~Ly, U.~Boehm, M.~Marsman, D.~S.
  Leslie, J.~J. Forster, E.-J. Wagenmakers, and H.~Steingroever (2017).
\newblock A tutorial on bridge sampling.
\newblock {\em Journal of Mathematical Psychology\/}~{\em 81}, 80--97.

\bibitem[\protect\citeauthoryear{Gronau, Singmann, and Wagenmakers}{Gronau
  et~al.}{2017}]{gronau2017bridgesampling}
Gronau, Q.~F., H.~Singmann, and E.-J. Wagenmakers (2017).
\newblock Bridgesampling: an {R} package for estimating normalizing constants.
\newblock {\em arXiv preprint arXiv:1710.08162\/}.

\bibitem[\protect\citeauthoryear{Hesselbo and Stinchcombe}{Hesselbo and
  Stinchcombe}{1995}]{hesselbo1995monte}
Hesselbo, B. and R.~B. Stinchcombe (1995).
\newblock Monte {C}arlo simulation and global optimization without parameters.
\newblock {\em Physical Review Letters\/}~{\em 74\/}(12), 2151--2155.

\bibitem[\protect\citeauthoryear{Hesterberg}{Hesterberg}{1995}]{hesterberg1995weighted}
Hesterberg, T. (1995).
\newblock Weighted average importance sampling and defensive mixture
  distributions.
\newblock {\em Technometrics\/}~{\em 37\/}(2), 185--194.

\bibitem[\protect\citeauthoryear{Jacob and Ryder}{Jacob and
  Ryder}{2014}]{jacob2014wang}
Jacob, P.~E. and R.~J. Ryder (2014).
\newblock The {W}ang--{L}andau algorithm reaches the flat histogram criterion
  in finite time.
\newblock {\em The Annals of Applied Probability\/}~{\em 24\/}(1), 34--53.

\bibitem[\protect\citeauthoryear{Kass and Raftery}{Kass and
  Raftery}{1995}]{kass1995bayes}
Kass, R.~E. and A.~E. Raftery (1995).
\newblock Bayes factors.
\newblock {\em Journal of the American Statistical Association\/}~{\em
  90\/}(430), 773--795.

\bibitem[\protect\citeauthoryear{Kiefer and Wolfowitz}{Kiefer and
  Wolfowitz}{1956}]{kiefer1956consistency}
Kiefer, J. and J.~Wolfowitz (1956).
\newblock Consistency of the maximum likelihood estimator in the presence of
  infinitely many incidental parameters.
\newblock {\em The Annals of Mathematical Statistics\/}~{\em 27}, 887--906.

\bibitem[\protect\citeauthoryear{Kong, McCullagh, Meng, and Nicolae}{Kong
  et~al.}{2006}]{kong2006}
Kong, A., P.~McCullagh, X.-L. Meng, and D.~Nicolae (2006).
\newblock Further explorations of likelihood theory for {M}onte {C}arlo
  integration.
\newblock In {\em Advances in Statistical Modeling and Inference: Essays in
  Honor of Kjell A. Doksum (Ed: V. Nair)}, pp.\  563--592. World Scientific
  Press.

\bibitem[\protect\citeauthoryear{Kong, McCullagh, Meng, Nicolae, and Tan}{Kong
  et~al.}{2003}]{kong2003theory}
Kong, A., P.~McCullagh, X.-L. Meng, D.~Nicolae, and Z.~Tan (2003).
\newblock A theory of statistical models for {M}onte {C}arlo integration (with
  discussions).
\newblock {\em Journal of the Royal Statistical Society: Series B (Statistical
  Methodology)\/}~{\em 65\/}(3), 585--604.

\bibitem[\protect\citeauthoryear{Kou, Zhou, Wong, et~al.}{Kou
  et~al.}{2006}]{kou2006equi}
Kou, S., Q.~Zhou, W.~H. Wong, et~al. (2006).
\newblock Equi-energy sampler with applications in statistical inference and
  statistical mechanics.
\newblock {\em The Annals of Statistics\/}~{\em 34\/}(4), 1581--1619.

\bibitem[\protect\citeauthoryear{Liang}{Liang}{2005}]{liang2005generalized}
Liang, F. (2005).
\newblock A generalized {W}ang--{L}andau algorithm for {M}onte {C}arlo
  computation.
\newblock {\em Journal of the American Statistical Association\/}~{\em
  100\/}(472), 1311--1327.

\bibitem[\protect\citeauthoryear{Liang, Liu, and Carroll}{Liang
  et~al.}{2007}]{liang2007stochastic}
Liang, F., C.~Liu, and R.~J. Carroll (2007).
\newblock Stochastic approximation in {M}onte {C}arlo computation.
\newblock {\em Journal of the American Statistical Association\/}~{\em
  102\/}(477), 305--320.

\bibitem[\protect\citeauthoryear{Liu, Liang, and Wong}{Liu
  et~al.}{2001}]{liu2001theory}
Liu, J.~S., F.~Liang, and W.~H. Wong (2001).
\newblock A theory for dynamic weighting in monte carlo computation.
\newblock {\em Journal of the American Statistical Association\/}~{\em
  96\/}(454), 561--573.

\bibitem[\protect\citeauthoryear{Martino, Elvira, Luengo, and Corander}{Martino
  et~al.}{2017}]{martino2017layered}
Martino, L., V.~Elvira, D.~Luengo, and J.~Corander (2017).
\newblock Layered adaptive importance sampling.
\newblock {\em Statistics and Computing\/}~{\em 27\/}(3), 599--623.

\bibitem[\protect\citeauthoryear{Meng}{Meng}{2005}]{MengXiao-Li2005CCSa}
Meng, X.-L. (2005).
\newblock Comment: Computation, survey and inference.
\newblock {\em Statistical Science\/}~{\em 20\/}(1), 21--28.

\bibitem[\protect\citeauthoryear{Meng and Schilling}{Meng and
  Schilling}{2002}]{meng2002warp}
Meng, X.-L. and S.~Schilling (2002).
\newblock Warp bridge sampling.
\newblock {\em Journal of Computational and Graphical Statistics\/}~{\em
  11\/}(3), 552--586.

\bibitem[\protect\citeauthoryear{Meng and Wong}{Meng and
  Wong}{1996}]{meng1996simulating}
Meng, X.-L. and W.~H. Wong (1996).
\newblock Simulating ratios of normalizing constants via a simple identity: {A}
  theoretical exploration.
\newblock {\em Statistica Sinica\/}~{\em 6\/}(4), 831--860.

\bibitem[\protect\citeauthoryear{Mira and Nicholls}{Mira and
  Nicholls}{2004}]{mira2004bridge}
Mira, A. and G.~Nicholls (2004).
\newblock Bridge estimation of the probability density at a point.
\newblock {\em Statistica Sinica\/}~{\em 14\/}(2), 603--612.

\bibitem[\protect\citeauthoryear{Owen and Zhou}{Owen and
  Zhou}{2000}]{owen2000safe}
Owen, A. and Y.~Zhou (2000).
\newblock Safe and effective importance sampling.
\newblock {\em Journal of the American Statistical Association\/}~{\em
  95\/}(449), 135--143.

\bibitem[\protect\citeauthoryear{Peel and McLachlan}{Peel and
  McLachlan}{2000}]{peel2000robust}
Peel, D. and G.~J. McLachlan (2000).
\newblock Robust mixture modelling using the {\it t}-distribution.
\newblock {\em Statistics and Computing\/}~{\em 10\/}(4), 339--348.

\bibitem[\protect\citeauthoryear{Romero}{Romero}{2003}]{romero2003two}
Romero, M. (2003).
\newblock {\em On two topics with no bridge: Bridge sampling with dependent
  draws and bias of the multiple imputation variance estimator}.
\newblock Ph.\ D. thesis, University of Chicago, Department of Statistics.

\bibitem[\protect\citeauthoryear{Shao and Ibrahim}{Shao and
  Ibrahim}{2000}]{shao2000monte}
Shao, Q.-M. and J.~G. Ibrahim (2000).
\newblock {\em Monte {C}arlo methods in {B}ayesian computation}.
\newblock Springer Series in Statistics, New York, NY.

\bibitem[\protect\citeauthoryear{Tan}{Tan}{2004}]{tan2004likelihood}
Tan, Z. (2004).
\newblock On a likelihood approach for {M}onte {C}arlo integration.
\newblock {\em Journal of the American Statistical Association\/}~{\em
  99\/}(468), 1027--1036.

\bibitem[\protect\citeauthoryear{Tan}{Tan}{2013}]{tan2013calibrated}
Tan, Z. (2013).
\newblock Calibrated path sampling and stepwise bridge sampling.
\newblock {\em Journal of Statistical Planning and Inference\/}~{\em 143\/}(4),
  675--690.

\bibitem[\protect\citeauthoryear{Veach and Guibas}{Veach and
  Guibas}{1995}]{veach1995optimally}
Veach, E. and L.~J. Guibas (1995).
\newblock Optimally combining sampling techniques for {M}onte {C}arlo
  rendering.
\newblock In {\em Proceedings of the 22nd Annual Conference on Computer
  Graphics and Interactive Techniques}, pp.\  419--428. ACM, New York, NY.

\bibitem[\protect\citeauthoryear{Villani}{Villani}{2003}]{villani2003topics}
Villani, C. (2003).
\newblock {\em Topics in optimal transportation}.
\newblock Number~58. American Mathematical Society, Providence, RI.

\bibitem[\protect\citeauthoryear{Voter}{Voter}{1985}]{voter1985monte}
Voter, A.~F. (1985).
\newblock A {M}onte {C}arlo method for determining free-energy differences and
  transition state theory rate constants.
\newblock {\em The Journal of Chemical Physics\/}~{\em 82\/}(4), 1890--1899.

\bibitem[\protect\citeauthoryear{Voter and Doll}{Voter and
  Doll}{1985}]{voter1985dynamical}
Voter, A.~F. and J.~D. Doll (1985).
\newblock Dynamical corrections to transition state theory for multistate
  systems: Surface self-diffusion in the rare-event regime.
\newblock {\em The Journal of Chemical Physics\/}~{\em 82\/}(1), 80--92.

\bibitem[\protect\citeauthoryear{Wang and Landau}{Wang and
  Landau}{2001}]{wang2001efficient}
Wang, F. and D.~Landau (2001).
\newblock Efficient, multiple-range random walk algorithm to calculate the
  density of states.
\newblock {\em Physical Review Letters\/}~{\em 86\/}(10), 2050--2053.

\bibitem[\protect\citeauthoryear{Wong and Liang}{Wong and
  Liang}{1997}]{wong1997dynamic}
Wong, W.~H. and F.~Liang (1997).
\newblock Dynamic weighting in monte carlo and optimization.
\newblock {\em Proceedings of the National Academy of Sciences\/}~{\em
  94\/}(26), 14220--14224.

\end{thebibliography}

\section*{Disclaimer}

This document is being distributed for informational and educational purposes only and is not an offer to sell or the solicitation of an offer to buy any securities or other
 instruments. The information contained herein is not intended to provide, and should not be relied upon for, investment advice.
 The views expressed herein are not necessarily the views of Two Sigma Investments, LP or any of its affiliates (collectively, “Two Sigma”).
 Such views reflect the assumptions of the author(s) of the document and are subject to change without notice. The document may employ data derived from third-party sources.
 No representation is made by Two Sigma as to the accuracy of such information and the use of such information in no way implies an endorsement of the source of such information
 or its validity.

The copyrights and/or trademarks in some of the images, logos or other material used herein may be owned by entities other than Two Sigma. If so, such copyrights and/or
trademarks are most likely owned by the entity that created the material and are used purely for identification and comment as fair use under international copyright
and/or trademark laws. Use of such image, copyright or trademark does not imply any association with such organization (or endorsement of such organization) by Two
Sigma, nor vice versa.

\appendix
\section*{\LARGE Supplementary Material: Appendices}
\section{Proof of Theorem 1}\label{app:proof}

Let $t(\theta,\omega) =\tr_\theta(\omega)$, then we can write   $\widetilde{W} =t({\Psi}, W)$ and
$\widetilde{X}=t(\Theta, X).$  Therefore,  $\tilde p$ and $\phi$ are related to
$p_{\text{\tiny$\Psi, \hspace{-0.06cm} W$}}$ and $\phi_{\text{\tiny$\Theta, \hspace{-0.06cm} X$}}$
of (\ref{equ::joint_dis_psi_w}) respectively, via the \textit{same} map $t: \Pi\times\Omega\rightarrow \Omega$.
Claim  (I) then follows from the   monotone property  of $f$-divergence \citep{ali1966general}:
\begin{equation*}\label{equ::main_inequality}
\mathcal{D}_f(\tilde{p}||\phi)\leqslant\mathcal{D}_f(p_{\text{\tiny$\Psi, \hspace{-0.06cm} W$}}||\phi_{\text{\tiny$\Theta, \hspace{-0.06cm} X$}} )= \mathcal{D}_f(p||\phi_{\mix}),
\end{equation*}
where the last equality holds because  $p_{\text{\tiny$\Psi, \hspace{-0.06cm} W$}}/\phi_{\text{\tiny$\Theta, \hspace{-0.06cm} X$}} =p/\phi_{\mix}$, a consequence of  (\ref{equ::joint_dis_psi_w}).

To prove (II), we use the fact from \citet{ali1966general} that,   when $f$ is strictly convex,   the inequality in (I)
becomes an equality if and only if $t$ is a sufficient statistic for the distribution family
$\{ p_{\text{\tiny$\Psi, \hspace{-0.06cm} W$}}, \phi_{\text{\tiny$\Theta, \hspace{-0.06cm} X$}}\}$.
By the well-known factorization theorem for sufficiency, the latter condition is the same as requiring that
$p_{\text{\tiny$\Psi, \hspace{-0.06cm} W$}}(\theta, \omega)/\phi_{\text{\tiny$\Theta, \hspace{-0.06cm} X$}}(\theta, \omega)$ depends on $(\theta, \omega)$ only through $t=t(\theta, \omega)= \tr_\theta(\omega)$, almost surely
with respect to $\measurev\times\measure$. But from (\ref{equ::joint_dis_psi_w}) and $\omega=\tr^{-1}_\theta(t)=\trh_\theta(t)$, we have
$$
p_{\text{\tiny$\Psi, \hspace{-0.06cm} W$}}(\theta, \omega)/\phi_{\text{\tiny$\Theta, \hspace{-0.06cm} X$}}(\theta, \omega)=p(\trh_{\theta}(t))/\phi_{\mix}(\trh_{\theta}(t))=\ell(\theta;t).
$$
Consequently,  $t$ is sufficient if and only if  $\ell(\theta;t)$ is  free of $\theta$, and hence (II).

%\section{EM details}\label{app:em}

\section{Setting Tuning Parameters: Guidance from Simulation Studies}\label{app:tuning}

We consider the tuning parameters for the estimator $\hat{\lambda}^{\pX}_{\text{\tiny H}} = \dfrac{1}{2}
\left(\hat{\lambda}^{\pX}_{\text{\tiny H$_1$}}+ \hat{\lambda}^{\pX}_{\text{\tiny H$_2$}}\right)$ introduced in Section \ref{adaptive_bias}, where $\mathcal{X}=U$ or ``mix".
As in Section \ref{sec:examples}, we drop the subscript $\text{H}$ (since we always use the estimation strategy in Section \ref{adaptive_bias}) and replace it by ``$\text{opt}$" to indicate that here we use the optimal bridge $\alpha\opt$ in (\ref{eq:opti}).
%In our algorithm for obtaining $\hat{\lambda}_\alpha^\pU$ or $\hat{\lambda}_\opt^\pmix$ t
 There are three tuning parameters:
\begin{itemize}
\item
$K$: the number of components in the normal mixture model $\phi_\mix(\cdot;\bz)$;\item
$L$: the number of draws from $p$ for estimating $\bz$,   with the restriction $L\leqslant n/2$;
\item
$m$: the number of draws from $N(0,I_D)$ or $\phi_\mix$.
\end{itemize}
%\subsection{Choosing Tuning Parameters}\label{tuning_paras_selection}
%Here, we use simulation results to demonstrate how
%the tuning parameters affect
%the statistical efficiency of $\hat{\lambda}_\opt^\pU$ and $\hat{\lambda}_\opt^\pmix$,
%and the associated computational costs.
By investigating the performance of the estimators $\hat{\lambda}_\opt^\pU=\dfrac{1}{2}\left(\hat{\lambda}_{\opt,1}^\pU+\hat{\lambda}_{\opt,2}^\pU\right)$ and
$\hat{\lambda}_\opt^\pmix=\dfrac{1}{2}\left(\hat{\lambda}_{\opt,1}^\pmix+\hat{\lambda}_{\opt,2}^\pmix\right)$
under different choices of $(K,L,m)$, we gain  practical guidance for choosing these tuning parameters.
To form a sensible compromise between  statistical and computational efficiency, we adopt
precision per CPU second ($PpS$), i.e.,  $(\Var \times \text{CPU seconds})^{-1}$ as our metric for evaluation.  We set the sample size $n$ to be $10^4$, which is sufficiently large to allow us to investigate the impact of large $K$ on estimator performance.
For each $(K,L,m)$, we generate
$10^4$ replicate data sets, each consisting of $m$ independent draws from $p$,  the 10-dimensional skewed-{\it t} mixture density introduced in Section \ref{adaptive_bias}.
If not specified, then
$L=\min\left(50K, n/2\right)$ and $m=n$.

\subsection{Impact of $K$}\label{impact_of_k_section}

\begin{figure}[t]
\begin{center}
\includegraphics[width=0.6\textwidth,trim=47mm 112mm 48mm 104mm,clip]{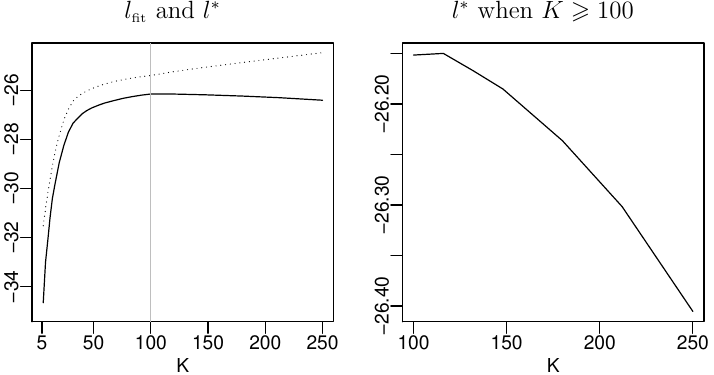}
\caption{\label{fig1_10D_max_likelihood}\small{(Dotted line) $\bar{l}_\fit$; (solid lines) $\bar{l}^*$. The gray vertical line
marks the value of $K$, around which $\bar{l}^*$ changes from increasing to decreasing as $K$ increases. The  right-hand panel shows $\bar{l}^*$ for $K$ ranging from 100 to 250 (but notice  the scale change on the vertical axis).
}}
\end{center}
\end{figure}

Figure~\ref{fig1_10D_max_likelihood} illustrates why
the variance and the RMSE of
$\hat{\lambda}_\opt^\pU$ in Figure \ref{bias_sd_trade_off} (Section \ref{specific_method}) decrease as $K$ increases.
The dotted line in Figure~\ref{fig1_10D_max_likelihood} is
the scaled  maximized log-likelihood $\bar{l}_{\fit}$, defined as
\begin{equation*}
\bar{l}_\fit = \frac{1}{L}\sum_{i=1}^L\log\left(\phi_\mix(w_i;\tilde{\bz}_\emsize)\right),
\end{equation*}
where $\tilde{\bz}_\emsize$ is the MLE of $\bz$ based on  $\{w_1,\ldots, w_\emsize\}$  and obtained via the EM algorithm.
The quantity  $\bar{l}_{\fit}$ measures how well  the calibrated
$\phi_\mix$ fits to the $L$ draws used for estimating $\bz$,
and is an increasing function of $K$.
The solid line in Figure \ref{fig1_10D_max_likelihood} represents the scaled log-likelihood $\bar{l}^*$ for the other half of the draws from $p$, i.e.,
\begin{equation*}
\bar{l}^* = \frac{2}{n}\sum_{i=n/2+1}^n\log\left(\phi_\mix(w_i;\tilde{\bz}_\emsize)\right).
\end{equation*}
The quantity  $\bar{l}^*$ provides an ``out of sample" assessment of the fit of $\phi_\mix$ to $p$. For moderate values of $K$, increases in $K$  (and hence increases in $\bar{l}_{\fit}$) generally result in increases in $\bar{l}^*$ because the mixture model better captures the mass of $p$.
%the resulting $\phi_\mix$ has more overlap with $p$.
%These improvements in fit also correspond to  statistical efficiency of $\hat{\lambda}_{\opt}^\pU$ increases as $K$ increases.

%We set $L=\min\left(50K, n/2\right)$, and will explain
%the choice of $L$ later.

\begin{figure}[t]
\begin{center}
 \renewcommand{\arraystretch}{0.28}
 \setlength{\tabcolsep}{2pt}
\includegraphics[width=0.9\textwidth,trim=27mm 115mm 24mm 108mm,clip]{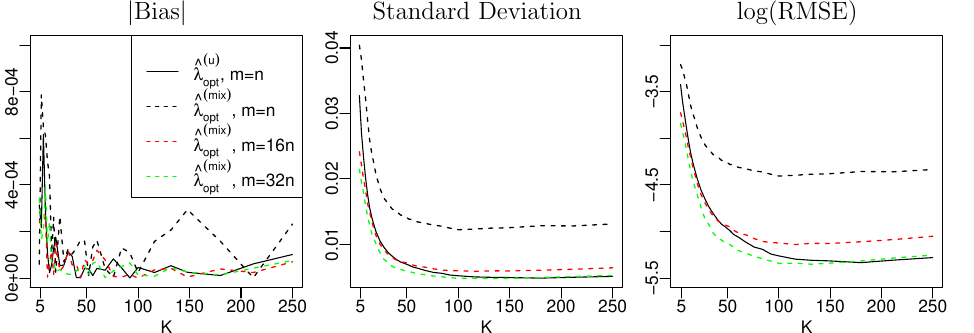}
  \caption{\label{10D_warpU_vs_mix_diff_K_diff_m}\small{The three columns show $|$bias$|$,
standard deviation and log(RMSE)  of
$\hat{\lambda}_\opt^\pU$ (solid lines) and $\hat{\lambda}_\opt^\pmix$ (dashed lines).
Different colors correspond to different values of $m$ in the estimators. Black: $m=n$;
Red: $m=16n$; Green: $m=32n$. }}
\end{center}
\end{figure}

However, for a large $K$,
the Normal  mixture model overfits the
$L\leqslant n/2$ draws used to estimate $\bz$.
% and with a finite sample,
%However, with a finite sample from $p$, overfitting of $\phi_\mix$ to the  will occur if $K$ is very large.
Figure~\ref{fig1_10D_max_likelihood} (right) shows that $\bar{l}^*$
decreases slightly when $K$ exceeds 100, indicating a slight increase in
the divergence between $p$ and $\phi_\mix(\cdot;\tilde{\bz}_\emsize)$.
Figure \ref{10D_warpU_vs_mix_diff_K_diff_m} shows
the $|$bias$|$, standard deviation, and RMSE (on a logarithmic scale)
of $\hat{\lambda}_\opt^\pU$ (solid lines) and $\hat{\lambda}_\opt^\pmix$ (dashed lines),
for $K$ ranging from 5 to 250.
When $K$ exceeds $n/100$, there is a slight increase in both the variance
and the RMSE of these estimators as $K$ continues to increase. Thus, the statistical efficiency of the estimators suffers when overfitting occurs as expected.
%Interestingly, $\hat{\lambda}_\opt^\pU$ appears to be less affected by overfitting than $\hat{\lambda}_\opt^\pmix$.

\begin{figure}[t]
\begin{center}
\includegraphics[width=0.6\textwidth,trim=47mm 113mm 48mm 102mm,clip]{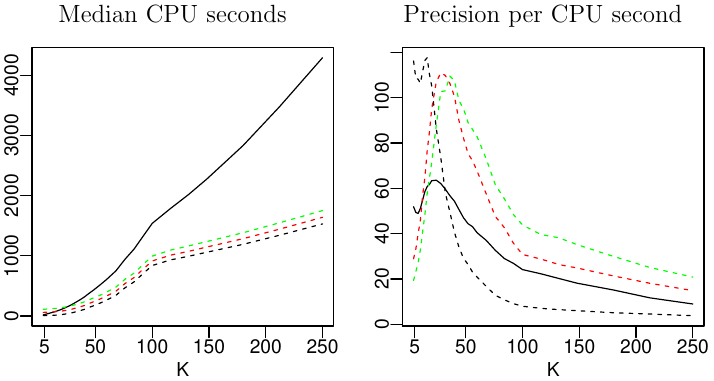}
\caption{\label{fig::10D_PpS_156}\small{The total computational cost $T^\pX_\opt$ (left),  and the precision per CPU second (right) of the
optimal bridge sampling estimators $\hat{\lambda}_{\opt}^\pU$ (solid lines, $m=n$) and
$\hat{\lambda}_{\opt}^\pmix$ (dashed lines) with $m=n$ (black), $16n$ (red), and $32n$ (green).
}}
\end{center}
\end{figure}

Figure \ref{fig::10D_PpS_156} (left) shows
the computational cost of
$\hat{\lambda}_\opt^\pU$ (solid line) and $\hat{\lambda}_\opt^\pmix$ (dashed line).
When $K>n/100$, the total number of CPU seconds $T_\opt^\pU$ exhibits  quadratic growth as a function of $K$, whereas
 $T_\opt^\pmix$  grows linearly with $K$. %(see Appendix \ref{computation} for explanation).
%Since there is little gain in statistical efficiency for $K>n/100$, the computational cost of choosing $K$ higher than $n/100$ cannot be justified.
Figure
\ref{fig::10D_PpS_156} (right) plots the $PpS$.
The largest $PpS$ is obtained when $K$ is between 20 and 30.
%For very small $K$, $PpS$
%increases with $K$, meaning that the reduction of variance
%is faster than the increase of computational costs, whereas when $K>30$,
%the curve of $PpS$ drops.


Based on the above simulations,
we suggest setting $K\leqslant n/100$
to avoid overfitting to the $L$ draws and unnecessary computational cost.
At this time, we have not been able to obtain any simple rule for constraining  the optimal $K$ further.
We want $K$ to be large enough to induce sufficient overlap between $\phi$ and $\tilde{p}$, but small enough to limit computational cost, both of which require problem specific knowledge.
However, as discussed in the main text, the beauty of Warp-U bridge sampling is that it performs very well even when $\phi_{\mix}$ is {\it not} a great fit to $p$.

%When $K$ is reasonably large, or when $t_q>>t_\phi$, which is often the case, $t_{\ttBS}^{\pU}\approx K t_{\ttBS}^{\pmix}$.
%Besides, the time required for Warp-U transformation is negligible compares with that for
%bridge sampling, that is $t_{\text{\tiny Tr}}^{\pU}<<t_{\ttBS}^{\pU}$. So
%approximately, $T^{\pU}\approx T_{\ttEM}+KT_{\ttBS}^{\pmix}$, meaning
%$T^{\pU} \geqslant T^{\pmix}$, meaning that

\subsection{Impact of $L$}\label{app:settingl}

Other factors being fixed, larger $L$ on average results in more overlap between $p$ and $\phi_{\mix}$, and hence between $\tilde{p}$ and $\phi$. Consequently, the statistical efficiency of $\hat{\lambda}^{\pU}_{\opt}$ increases with $L$.
It follows that if computational cost is not a concern,
we should use all of the draws in one half of the draws from $p$ to estimate $\bz$,
and to apply the corresponding Warp-U transformation to the other half
in order
to obtain $\hat{\lambda}_{\opt,i}^{\pU}$, for $i=1,2$.
 However,  increasing $L$ above moderate values may yield diminishing returns, even in the case where $p$ is exactly a Normal mixture so that the $f$-divergence between $p$ and $\phi_{\mix}(\cdot;\widetilde{\bz}_{\emsize})$  tends to zero as $L \rightarrow \infty$.

\begin{figure}[t]
 \renewcommand{\arraystretch}{0.28}
 \setlength{\tabcolsep}{2pt}
\begin{center}
\includegraphics[width=0.9\textwidth,trim=27mm 115mm 24mm 106mm,clip]{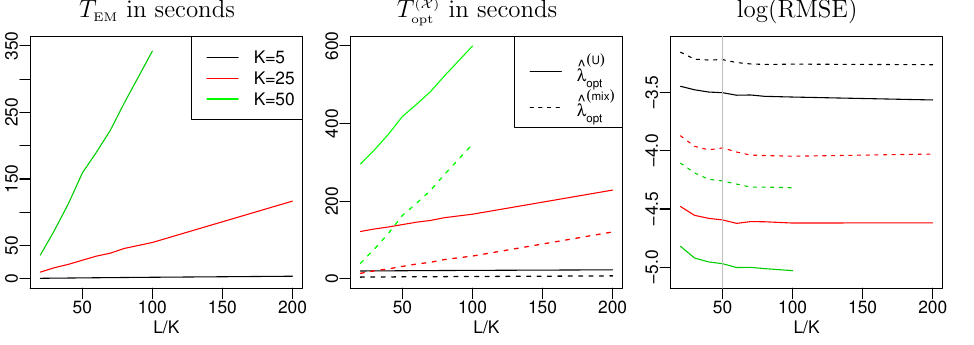}
\caption{\label{fig::10D_L_over_K}\small{The impact of $L$ on $T_\ttEM$ (left), $T_\opt^\pX$ (middle), and the log RMSE of
$\hat{\lambda}_\opt^\pX$ (right). Black lines: $K=5$; red lines: $K=25$; green lines: $K=50$.
When $K=50$, we can  take $L/K$ only up to 100, because of the restriction that $L\leqslant n/2$.
}}
\end{center}
\end{figure}

Figure \ref{fig::10D_L_over_K} shows
the impact of $L$ on
(i) $T_\ttEM$,
 the computational cost in CPU seconds of estimating $\bz$;
 (ii)   $T_\opt^\pU$ and $T_\opt^\pmix$,
the total computational cost in CPU seconds of obtaining $\hat{\lambda}_\opt^\pU$ and $\hat{\lambda}_\opt^\pmix$;
and (iii) the RMSE.
%The value of $L$ has similar impact on the variance and computational costs as $K$.
Intuitively, we should increase the size of the sample for estimating $\bz$ linearly with
$K$, so we compare estimators with different values of $L/K$.
%In terms of the computational costs,
In the left panel of Figure \ref{fig::10D_L_over_K} we can see that $T_\ttEM$ increases linearly with $L$, for fixed $K$, and therefore so does $T_\opt^\pX$ (middle  panel), because $L$ affects only the EM part.
%, as shown in Figure \ref{fig::10D_L_over_K} (left and middle),
%where $K$ is set to be 5 (black lines), 25 (red lines), and 50 (green lines).
%, whereas small $L/K$
%may cause an overfitting problem and thus
%large divergence between $p$ and $\phi_\mix(\cdot; \tbz)$.
The right panel of Figure \ref{fig::10D_L_over_K} shows that RMSE decreases as $L/K$ increases, but
the reduction rate becomes very small when $L/K>50$.
We obtained similar results (not shown) for the 50-dimension example in Section \ref{sec:examples}.
%impact of $L/K$ on the RMSE of $\hat{\lambda}_\opt^\pX$
  Thus, a reasonable rule of thumb  is  to set $L=\min(50K,n/2)$.

%Small $L$ will result in unstable estimation of
%$\bz$ and thus large RMSE of the estimator. However,  in this section,
%we show in simulation that
%when $L\geqslant 50K$, especially when $L\geqslant 100K$, increasing $L$ has very little improvement on the estimator, but requires expensive computation
%resources. Thus, we propose to set $L$ to be between $50K$ and $100K$.

\subsection{Impact of $m$ and a Comparison of $\hat{\lambda}_\opt^\pmix$ and $\hat{\lambda}_\opt^\pU$}
As with $K$ and $L/K$, larger
$m$ improves the precision of the estimators but
also increases the computational cost. In the case of $\hat{\lambda}_\opt^\pmix$, setting $m>n$ may only increase the computational cost by a small amount and therefore can be a worthwhile trade-off. In particular,
 when $K$ is large and target evaluations are inexpensive,
$T_\ttEM$ will typically dominate $T_\ttBS^\pmix$ meaning that the additional computational cost introduced by increasing $m$ is small and likely acceptable given  the potential for substantial improvements in statistical efficiency.
%For easy reference, we use
Indeed, referring back to Figure \ref{fig::10D_PpS_156} (left), when $K$ is moderate or large, we see that
the difference in the computational cost of $\hat{\lambda}_\opt^\pmix$ across the cases $m=n$ (black dashed line), $16n$ (red  dashed line), and $32n$ (green  dashed line),
is negligible compared with the total cost $T_\ttEM + T_\ttBS^\pmix$.
%The variance of $\hat{\lambda}_\opt^\pmix$, however,  drops substantially when $m$ increases from $n$ to $32n$ in Figure \ref{10D_warpU_vs_mix_diff_K_diff_m} (middle).
%by nearly 80\% when $m$ increases from $n$ to $
 %$\hat{\lambda}_\opt^\pmix(32n)$ is only 20\% of that of $\hat{\lambda}_\opt^\pmix(n)$, as
%shown in  and \ref{fig::10D_diff_m_over_n}.
%In fact, $\hat{\lambda}_\opt^\pmix(16n)$ and $\hat{\lambda}_\opt^\pmix(32n)$ are comparable with  $\hat{\lambda}_\opt^\pU(n)$ in terms of statistical efficiency, but $\hat{\lambda}_\opt^\pU(n)$ is much more computationally costly.
%Consequently,
Consequently, $\hat{\lambda}_\opt^\pmix(16n)$
and $\hat{\lambda}_\opt^\pmix(32n)$ have larger $PpS$ than $\hat{\lambda}_\opt^\pU(n)$
for moderate and large $K$, where $\hat{\lambda}_{\opt}^\pX(m)$ denotes
the estimator $\hat{\lambda}_{\opt}^\pX$ with a specific setting of $m$; see Figure \ref{fig::10D_PpS_156} (right).

%the
%statistical efficiency of $\hat{\lambda}_\opt^\pU$ can also be improved by increasing $m$,
%as shown in  (middle), where the solid lines
%represent the standard deviation of $\hat{\lambda}_\opt^\pU$ with different values of
%$m$.
%but with significantly increased computational cost.
%Hence, in most cases,

\begin{figure}[t]
\begin{center}
 \renewcommand{\arraystretch}{0.28}
 \setlength{\tabcolsep}{2pt}
\includegraphics[width=0.9\textwidth,trim=27mm 114mm 24mm 108mm,clip]{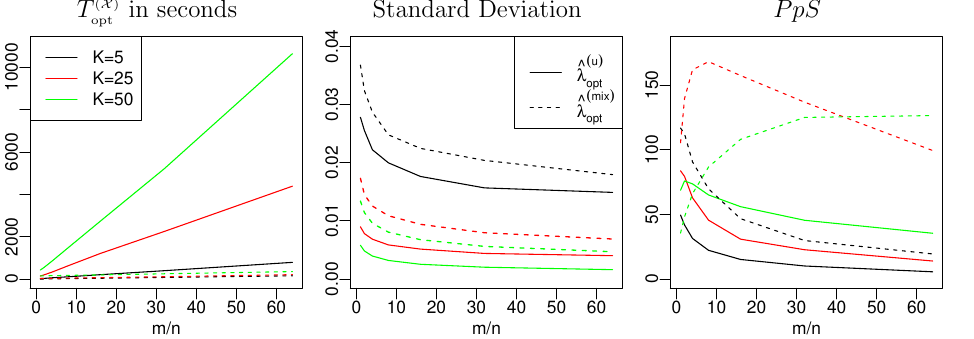}
\caption{\label{fig::10D_diff_m_over_n}\small{The total CPU time $T_\opt^\pX$ (left),
the standard deviation (middle), and the $PpS$ of $\hat{\lambda}_{\opt}^\pU$ (solid lines) and $\hat{\lambda}_{\opt}^\pmix$ (dashed lines)
with various choices  of $m$ for $K=5$ (black),
25 (red), and 50 (green). }}
\end{center}
\end{figure}

\begin{comment}
Consistent with our theoretical results,
Figures \ref{10D_warpU_vs_mix_diff_K_diff_m},
\ref{fig::10D_L_over_K}, and
\ref{fig::10D_diff_m_over_n}
all illustrate that, for the same $(K,L,m)$,
$\hat{\lambda}_\opt^{\pU}$ has better statistical efficiency than
$\hat{\lambda}_\opt^{\pmix}$,
but $\hat{\lambda}_\opt^{\pU}$ is computationally much more costly than $\hat{\lambda}_\opt^{\pmix}$.
Figure \ref{10D_warpU_vs_mix_diff_K_diff_m} (middle)
shows that the difference between the variances of
$\hat{\lambda}_\opt^\pU$
and $\hat{\lambda}_\opt^\pmix$ sometimes increases
as $K$ increases. A possible explanation is as follows.
In the Warp-U transformation,
the overlap of $\tilde{p}^{\text{\tiny (k)}}$ and $\tilde{\phi}^{\text{\tiny (k)}}$
remains the same as that of $p^{\text{\tiny (k)}}$ and $\phi^{\text{\tiny (k)}}$, and
the additional overlap comes from  rematching
$\tilde{p}^{\text{\tiny (k)}}$ with the remainder of $\tilde{\phi}^{\text{\tiny (j)}}$ (for $j\neq k$)
that does not already overlap with $\tilde{p}^{\text{\tiny (j)}}$.
The total number of possible rematching pairs  is $K(K-1)/2$, so
as $K$ increases, there is more opportunity for the aforementioned cross-overlap overlap to occur (although the .
%This may also be the reason why $\hat{\lambda}_\alpha^\pU$ is less affect by overfitting than $\hat{\lambda}_\alpha^\pmix$.
\end{comment}

Figure \ref{fig::10D_diff_m_over_n} further confirms these findings, and demonstrates that increasing $m$ does not offer the same improvements in $PpS$ in the case of $\hat{\lambda}_\opt^\pU$.  Figure \ref{fig::10D_diff_m_over_n} (left) shows that
both $T_\opt^\pU$ (solid lines) and $T_\opt^\pmix$ (dashed lines)
%that the total computational costs
%of $\hat{\lambda}_\opt^\pU$ (solid lines) and $\hat{\lambda}_\opt^\pmix$ (dashed lines)
grow linearly
as $m$ increases from $n$ to $64n$.
Figure \ref{fig::10D_diff_m_over_n} (middle) shows that
the standard deviation
of $\hat{\lambda}_\opt^\pX$ is inversely related to $m$. Figure \ref{fig::10D_diff_m_over_n} (right) shows that the $PpS$ of $\hat{\lambda}_\opt^\pU$ usually  decreases as $m$ increases, which is a result of the high computational cost associated with increases in $m$. However, we again see that $m > n$ can improve the $PpS$ for $\hat{\lambda}_\opt^\pmix$, because  the computational cost of the bridge sampling step, $T_\ttBS^\pmix$, is low compared with $T_\ttEM$ and $T_\ttBS^\pU$. %, as already mentioned.

Despite the above findings, given a fixed sample of size $n$ from $p$,
the best statistical efficiency achieved by $\hat{\lambda}_\opt^\pU$
is better than that of $\hat{\lambda}_\opt^\pmix$. The higher computational cost of  $\hat{\lambda}_\opt^\pU$  can  be mitigated  by parallel computing, for instance, which is a topic for further research. Moreover, in cases where target evaluations are expensive, there will be much less opportunity for improving the $PpS$ of $\hat{\lambda}_\opt^\pmix$ by increasing $m$ because $T_\ttEM$ will not dominate $T_\ttBS^\pmix$. In these cases the superiority of  $\hat{\lambda}_\opt^\pmix$ in terms of $PpS$  disappears, as confirmed by the example in Section \ref{sec:comparison}.

To sum up, we recommend using $L=50K$ draws from $p$ to estimate $\bz$. The variance of $\hat{\lambda}_\opt^\pX$ can be effectively reduced by
increasing $K$ and/or $m$ up to certain levels.
The rates of reduction in variance are different for $K$ and $m$.
When $K$ is small, increasing $K$ reduces the variance faster than
increasing $m$; when $K$ is large, increasing $m$ is more beneficial for reducing  the
variance. For the estimator $\hat{\lambda}_\opt^\pmix$, having a large $m$, e.g., $m=10n$,
is recommended
in cases where  $T_\ttBS^\pmix$ is low compared with $T_\ttEM$.

\section{Efficiency of Diagonal vs. Full Covariance Matrices}\label{app:cov}

\begin{figure}
\begin{center}
\includegraphics[width=0.9\textwidth,trim=29mm 119mm 24mm 108mm,clip]{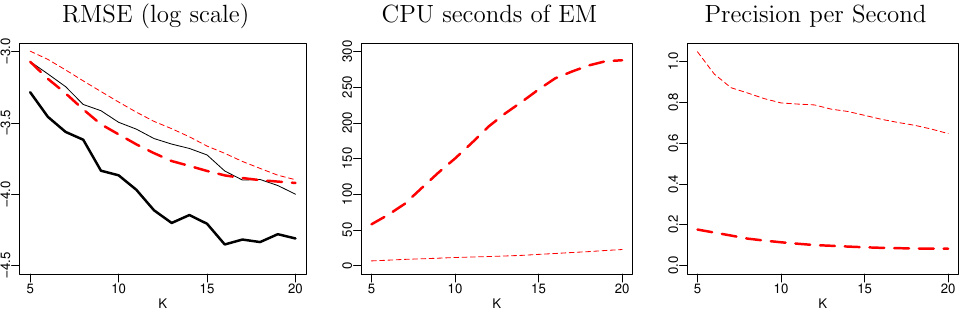}
\caption{\label{bias_sd_trade_off7}\small{(Left) Log RMSE of different estimators; (middle)
CPU seconds used by the EM algorithm; (right) precision per CPU second defined as (1/MSE)/time.
(Solid lines) Warp-U bridge sampling with $\widetilde{\bz}_\text{\tiny $\mathcal{I}$}$;
 (dashed lines)
the average of the two Warp-U bridge sampling estimators with
half of the draws fro $p$ used for estimating $\bz$ and the other half for bridge sampling. (Thin lines)
diagonal covariance matrices;
(thick lines) full covariance matrices.}}
\end{center}
\end{figure}

Here we compare the computational and statistical efficiency of using diagonal and full covariance matrices in $\phi_\mix$ in (\ref{equ::mixture_density_general}). The left panel of Figure \ref{bias_sd_trade_off7} shows
the log RMSE of $\hat{\lambda}_\text{\tiny $\mathcal{I}$,Diag}^{\pU}$,
$\hat{\lambda}_\text{\tiny $\mathcal{I}$,Full}^{\pU}$,
$\hat{\lambda}_\text{\tiny H,Diag}^{\pU}$, and
$\hat{\lambda}_\text{\tiny H,Full}^{\pU}$.
On average, the log RMSE of $\hat{\lambda}_\text{\tiny $\mathcal{I}$,Diag}^{\pU}$ (thin solid line) is about 50\%  larger than
that of $\hat{\lambda}_\text{\tiny $\mathcal{I}$,Full}^{\pU}$ (thick solid line). Recall that the $\mathcal{I}$ subscript indicates that an additional independent set of draws from $p$ is used to estimate the covariance matrices, and hence there is no adaptive bias and also more degrees of freedom to estimate the covariance matrices. In such cases, there is a benefit to using the full covariance matrices, because they are accurately estimated.  However, when we need to estimate the covariance matrices from the original set of draws from $p$ there is not much advantage to using full covariance matrices; the log RMSE
 of $\hat{\lambda}_\text{\tiny H,Diag}^{\pU}$ (thin dashed line) is only 16.7\%
larger than that of
$\hat{\lambda}_\text{\tiny H,Full}^{\pU}$ (thick dashed line), and the difference
diminishes as $K$ increases.
This is because, when $K$ is large,
over-fitting becomes more serious for
the full-matrix model
due to the additional
$KD(D-1)/2$ parameters in the model. The diagonal-matrix model,
being much more parsimonious,  continues to fit the date (i.e., the draws from $p$) well and
the resulting log RMSE decreases at a stable rate.
%
%than the
%diagonal-covariance-matrix model,
%
%the resulting model will overfit the data
%
%we need to estimate
% when $\scale_k$ are not restricted to diagonal matrices.
%With only $n/2$ samples used for the estimation,
%, so
%the RMSE of $\hat{\lambda}_\text{\tiny H,Full}^{\pU}$ gradually flatten out.
%On the other hand, the Gaussian mixture model with diagonal covariance matrices
%is very different from $p$, so as $K$ increases, the model continues to fit the data better and
%the RMSE decreases at a stable rate.
%Second, when $K$ approaches to 20, the Gaussian mixture model with full covariance matrices fits the data better than
%that with diagonal matrices. So the thin dashed line decreases at a relative stable rate as $K$ increases, whereas the
%thick dashed line gradually flatten out.
%When $K$ is large, we may not have enough data to estimate $\bz$ and the resulted model may overfit the data.

Figure \ref{bias_sd_trade_off7} (right) shows the CPU seconds for estimating $\bz$
via  EM.
On average in this study,  it takes 12 times longer to obtain $\phi_{\mix}$ with
full covariance matrices than with diagonal covariance matrices,  and the difference  increases
with the dimensionality.
In addition, in the bridge sampling step,  evaluating $\phi_{\mix}$
with full covariance matrices is much more costly than with diagonal covariance matrices. Therefore,
a small loss of statistical efficiency but  huge gain in  computational efficiency   justifies  the use of diagonal covariance matrices. Indeed, for reducing  RMSE, increasing  the number of mixture components $K$ seems to be more effective than using full covariance matrices. This is consistent with our intuition that, for the purposes of increasing distributional overlap,  it is more important to increase the chance that our model $\phi_\mix$  shares major modes with $p$ than to refine the curvature of $\phi_\mix$ at the modes.

%\section{GWL algorithm settings}\label{app:gwl}

\section{Details of GWL implementation}
\label{app:gwl_details}

A complication in specifying the GWL algorithm is identifying a suitable proposal distribution for the Metropolis-Hastings update in its Step 2(i) detailed in Section \ref{sec:gwl}. We reproduced the results for the 2 dimensional examples in \citet{liang2005generalized}, but found  it difficult to identify a simple proposal that results in comparable performance of the GWL algorithm in our 10 dimensional example. Perhaps a sophisticated method is needed to choose the proposal distribution in examples with more than a few dimensions. For our simulation study, we can avoid this issue by deliberately biasing our comparison in favor of the GWL algorithm (and hence making it harder to show the benefit of our proposal). That is, we adopted the true target density $p$ (which is distinct from the limiting distribution of the GWL algorithm) as a component of the proposal. More precisely, our proposal density is $\frac{40}{41} p_S(x)+ \frac{1}{41}u(x)$, where $p_S$ is $p$ restricted to $S=\bigcup_{i=2}^rS_i=[-20,20]^{10}$ and $u(x)$ is the uniform density on the additional subregion $S_{1}=[20,21]\times[-20,20]^9$. Using this proposal does not result in sampling directly from $p$ (or $p_S$) because the GWL algorithm attempts to sample energy bins uniformly. We verified that the performance of the GWL algorithm is substantially better with this ideal proposal density than under all other proposals that we considered in our investigations (mostly Normal densities with different variances).

The final part of the setup is to define the energy bins $E_1,\dots,E_{r}$, and we again do this in a way that is favorable to the GWL algorithm. We generate 1000 samples from the (ideal) proposal and compute their energies and also evaluate the energy at the true location parameter of each of the 25 skewed-{\it t} densities. We then define the energy bins to span the range of these energies in steps of $0.1$ (all energies outside this range are put in the end bins). This typically yields around 350 bins but the exact number varies with each run of the algorithm. We briefly experimented with different energy steps but found that $0.1$ gave the best results.

\end{document}